\documentclass[12pt,a4paper]{article}
\usepackage{color,graphics,amsmath,epsfig,rotating}

\begin{document}

\setlength{\unitlength}{1mm}


\def\ma{m_A}
\def\mhf{m_{1/2}}
\def\m0{m_0}
\def\ra{\rightarrow}
\def\neuto {\tilde\chi_1^0}
\def\mneuto{m_{\tilde{\chi}_1^0}}
\def\stauo{\tilde\tau_1}
\def\staur{\tilde{\tau_R}}
\def\ser{\tilde{e_R}}
\def\smur{\tilde{\mu}_R}
\def\mt{m_t}
\def\mbmb{m_b(m_b)}
\def\mslr{m_{\tilde l_R}}

\def\nn              {\notag}
\def\bce             {\begin{center}}
\def\ece             {\end{center}}

\def\mbf             {\boldmath}

\def\ti              {\tilde}

\def\a               {\alpha}
\def\b               {\beta}
\def\d               {\delta}
\def\D               {\Delta}
\def\g               {\gamma}
\def\G               {\Gamma}
\def\l               {\lambda}
\def\t               {\theta}
\def\s               {\sigma}
\def\S               {\Sigma}
\def\x               {\chi}

\def\sq              {\ti q}
\def\sqL             {\ti q_L^{}}
\def\sqR             {\ti q_R^{}}
\def\sf              {\ti f}

\def\st              {\ti t}
\def\sb              {\ti b}
\def\stau            {\ti \tau}
\def\snu             {\ti \nu}

\def\sqbar  {\Bar{\Tilde q}^{}}
\def\stbar  {{\Bar{\Tilde t}}}
\def\sbbar  {{\Bar{\Tilde b}}}

\def\Pp  {{\cal P}_{\!+}}
\def\Pm  {{\cal P}_{\!-}}

\def\ch              {\ti \x^\pm}
\def\chp             {\ti \x^+}
\def\chm             {\ti \x^-}
\def\nt              {\ti \x^0}

\newcommand{\msq}[1]   {m_{\ti q_{#1}}}
\newcommand{\msf}[1]   {m_{\ti f_{#1}}}
\newcommand{\mst}[1]   {m_{\ti t_{#1}}}
\newcommand{\msb}[1]   {m_{\ti b_{#1}}}
\newcommand{\mstau}[1] {m_{\ti\tau_{#1}}}
\newcommand{\mch}[1]   {m_{\ti \x^\pm_{#1}}}
\newcommand{\mnt}[1]   {m_{\ti \x^0_{#1}}}
\newcommand{\mq}{\mbox{$m_{\tilde{q}}$}}
\newcommand{\mhp}      {m_{H^+}}
\newcommand{\msg}      {m_{\ti g}}

\def\sg              {{\ti g}}
\def\msg             {m_{\sg}}

\def\tW              {\t_{\scriptscriptstyle W}}
\def\tsq             {\t_{\sq}}
\def\tst             {\t_{\st}}
\def\tsb             {\t_{\sb}}
\def\tstau           {\t_{\stau}}
\def\tsf             {\t_{\sf}}
\def\sth             {\sin\t}
\def\cth             {\cos\t}
\def\cst             {\cos\t_{\st}}
\def\csb             {\cos\t_{\sb}}

\def\onehf           {{\textstyle \frac{1}{2}}}
\def\oneth           {{\textstyle \frac{1}{3}}}
\def\twoth           {{\textstyle \frac{2}{3}}}

\def\rzw             {\sqrt{2}}

\def\BR              {{\rm BR}}
\def\mev             {{\rm MeV}}
\def\gev             {{\rm GeV}}
\def\tev             {{\rm TeV}}
\def\fb              {{\rm fb}}
\def\fbi             {{\rm fb}^{-1}}

\def\over            {\overline}
\def\MSbar           {{\overline{\rm MS}}}
\def\DRbar           {{\overline{\rm DR}}}
\def\DR              {{\rm\overline{DR}}}
\def\MS              {{\rm\overline{MS}}}

\def\isajet          {{\tt ISAJET\,7.71}}
\def\softsusy        {{\tt SOFTSUSY\,1.9}}
\def\spheno          {{\tt SPHENO\,2.2.2}}
\def\suspect         {{\tt SUSPECT\,2.3}}
\def\newsuspect      {{\tt SUSPECT\,2.3.4}}
\def\micromegas      {{\tt micrOMEGAs\,1.3.2}}

\def\isajetnn        {{\tt ISAJET}}
\def\softsusynn      {{\tt SOFTSUSY}}
\def\sphenonn        {{\tt SPHENO}}
\def\suspectnn       {{\tt SUSPECT}}

\newcommand{\beqn}{\begin{eqnarray}}
\newcommand{\eeqn}{\end{eqnarray}}

\newcommand{\eq}[1]  {\mbox{(\ref{eq:#1})}}
\newcommand{\fig}[1] {Fig.~\ref{fig:#1}}
\newcommand{\Fig}[1] {Figure~\ref{fig:#1}}
\newcommand{\tab}[1] {Table~\ref{tab:#1}}
\newcommand{\Tab}[1] {Table~\ref{tab:#1}}

\newcommand{\gsim}{\;\raisebox{-0.9ex}
           {$\textstyle\stackrel{\textstyle >}{\sim}$}\;}

\newcommand{\lsim}{\;\raisebox{-0.9ex}{$\textstyle\stackrel{\textstyle<}
           {\sim}$}\;}

\newcommand{\smaf}[2] {{\textstyle \frac{#1}{#2} }}


\definecolor{red}{rgb}{1,0,0}
\definecolor{lred}{rgb}{1,0.15,0}
\definecolor{dred}{rgb}{0.7,0,0}
\definecolor{ddred}{rgb}{0.5,0,0}

\definecolor{green}{rgb}{0,1,0}
\definecolor{lgreen}{rgb}{0.3,1,0}
\definecolor{dgreen}{rgb}{0,0.7,0}

\definecolor{blue}{rgb}{0,0,1}
\definecolor{lblue}{rgb}{0,0.5,1}
\definecolor{llblue}{rgb}{0,0.8,1}
\definecolor{dblue}{rgb}{0,0,0.7}
\definecolor{ddblue}{rgb}{0,0,0.5}

\definecolor{orange}{rgb}{1,0.5,0}
\definecolor{dorange}{rgb}{0.8,0.3,0}

\definecolor{pink}{rgb}{1,0,1}
\definecolor{lila}{rgb}{0.7,0,0.7}
\definecolor{purple}{rgb}{0.7,0,0.7}

\definecolor{gray}{rgb}{0.2,0.2,0.2}

\newcommand{\bb} {\color{blue}}
\newcommand{\bbf} {\color{blue}\bf}
\newcommand{\change} {\bf\mbf}


\begin{flushright}
   \vspace*{-18mm}
   LAPTH-1079/04\\
   CERN-PH-TH/2004-253
\end{flushright}
\vspace*{2mm}

\bce

{\Large\bf  Comparison of SUSY spectrum calculations and\\[3mm]
            impact on the relic density constraints from WMAP } \\[10mm]

{\large   G.~B\'elanger$^1$, S.~Kraml$^2$, A.~Pukhov$^3$}\\[4mm]

{\it 1) Laboratoire de Physique Th\'eorique LAPTH, F-74941 Annecy-le-Vieux, France\\
     2) CERN, Dep. of Physics, Theory Division, CH-1211 Geneva 23, Switzerland\\
     3) Skobeltsyn Inst. of Nuclear Physics, Moscow State Univ., Moscow 119992, Russia  }\\[4mm]


\ece

\begin{abstract}
We compare results of four public supersymmetric (SUSY) spectrum codes,
\isajet, \softsusy, \spheno\ and \suspect\ to estimate the present-day 
uncertainty in the calculation of the relic density of dark matter in mSUGRA 
models. We find that even for mass differences of about 1\% the spread in the 
obtained relic densities can be 10\%. In difficult regions of the parameter 
space, such as large $\tan\beta$ or large $\m0$, discrepancies in the relic 
density are much larger. 
We also find important differences in the stau co-annihilation region. 
We show the impact of these uncertainties on the bounds from WMAP for several 
scenarios, concentrating on the regions of parameter space most relevant for 
collider phenomenology. We also discuss the case of $A_0\not=0$ 
and the stop co-annihilation region. 
Moreover, we present a web application for the online
comparison of the spectrum codes.
\end{abstract}

\section{Introduction}

Since the extremely precise measurement of the cosmic microwave background 
by the WMAP experiment
\cite{Bennett:2003bz,Spergel:2003cb}, cosmology  has been used to
severely constrain models with cold dark matter candidates. The
prime example are supersymmetric models with R-parity conservation
where the neutralino LSP (lightest supersymmetric particle) 
is the cold dark matter
(see Ref.~\cite{Feng:2003zu} for a review of SUSY cosmology).
Requiring that the model provide the right amount of cold dark matter
\begin{equation}
   0.0945 \leq \Omega_{CDM} h^2 \leq 0.1287
\label{eq:wmap}
\end{equation}
at $2\sigma$ puts strong constraints on the parameter space of
the model, in particular in the mSUGRA scenario
\cite{Ellis:2003cw,Baer:2003yh,Chattopadhyay:2003xi,Profumo:2004at,
Baltz:2004aw,Belanger:2004ag,Lahanas:2003yz,Ellis:2004tc}.
Effectively, the relic density of dark matter imposes some
very specific relations among the parameters of the model.
Naturally, the question arises how precisely
$\Omega$\,\footnote{In what follows, $\Omega\equiv\Omega_{CDM}h^2$.}
is calculated in a supersymmetric model.
%
We therefore revisit the constraints from WMAP in the mSUGRA scenario 
taking into account uncertainties originating from the computation 
of the SUSY spectrum. 
In the standard approach, the relic density is $\Omega\propto
1/\langle\sigma v\rangle$, where $\langle\sigma v\rangle$ is the
thermally averaged cross section times the relative velocity of
the LSP pair.  This thermally averaged effective annihilation
cross section includes a sum over all annihilation channels for
the LSP as well as co-annihilation channels involving sparticles
that are close in mass to the LSP. The relic density then depends
on all the parameters of the MSSM (i.e.\ masses and couplings)
that enter the different annihilation/co-annihilation channels.
To calculate the relevant cross sections within the context of
a model defined at a high scale, say the GUT scale, one first needs
to solve the renormalization group equations to obtain the MSSM
parameters at the SUSY scale. Second, higher-order corrections to
the masses and couplings need to be calculated.
Many public or private spectrum calculators perfom this task.
The results are then used to calculate in an improved tree-level
approximation the effective annihilation cross-section of neutralinos
and the relic density of dark matter.
%
This kind of top-down approach is also the typical 
method to test high-scale models at the LHC \cite{Allanach:2004ud}.
To address the issue of the precision of the relic density computation
in mSUGRA, in this note
we compare the results of four public spectrum  codes,
\isajet~\cite{Paige:2003mg}, \softsusy~\cite{Allanach:2001kg},
\spheno~\cite{Porod:2003um} and \suspect~\cite{Djouadi:2002ze}, 
linking them to \micromegas~\cite{Belanger:2004yn} to compute $\Omega$.
Since  three of these codes, \isajet, \softsusy\ and \spheno, 
are of a comparable level as what concerns radiative corrections, 
the differences in their results seem to be a
good estimate of the present uncertainties due to
higher-order loop effects.
We also include \suspect\ in the discussion because it is a widely 
used program. However, since in contrast to the other three codes,  
\suspect\ has only 1-loop renormalization group (RG) running for the 
squark and slepton mass parameters we do not use it for the estimate 
of uncertainties.
Within a given program, one can also estimate the theoretical
uncertainty by, for example, varying the scale $M_{SUSY}$ at
which the electroweak-symmetry breaking conditions are imposed
and the sparticle masses are calculated.
This was discussed in Ref.~\cite{Allanach:2004xn} and uncertainties
on the relic density up to 20\% were found.

The  MSSM parameters that enter  the effective annihilation cross
section for the LSP  include all the ones contributing to the
annihilation and co-annihilation processes.
The relic density can then
depend on a large number of parameters. However,
because one needs, at least within the context of SUGRA models,
very specific mechanisms to satisfy the tight upper bound of WMAP,
only a few parameters are critical within each scenario \cite{Allanach:2004xn}.
Any shift in one of the critical variable can have a large impact
on the value of the relic density.
Within mSUGRA, the preferred scenarios are the $\stau$ co-annihilation,
the rapid Higgs annihilation and the higgsino-LSP scenarios.
The main channels are annihilation of neutralinos into fermion
pairs via s-channel Z or Higgs exchange, or via t-channel
sfermion exchange, as well as co-annihilation with sleptons.
For example, in the co-annihilation region, co-annihilation processes 
are suppressed by a factor $exp ^{-\Delta M/T_f}$ where $\Delta M$ is 
the mass difference with the LSP and
$T_f\approx \mneuto/25$ is the decoupling temperature.
Then it is the mass difference between the NLSP  
and LSP that introduces the
largest uncertainty in the prediction of the relic density.
In Ref.~\cite{Allanach:2004xn} it was shown that a 1~GeV correction
to the mass difference could lead to 10\% correction on the relic density.
In \cite{Allanach:2003jw} it has been pointed out that typical differences in 
the masses obtained by the spectrum calculator codes are of ${\cal O}(1\%)$,
large enough for the computational uncertainty to exceed the experimental
one of WMAP.
In other scenarios, the ones where annihilation proceeds through s-channel 
Z or Higgs exchange,  the  important parameters are the coupling of 
neutralinos to the Z or Higgs and the mass of the LSP in relation with 
the mass of the resonance, in general the mass of the pseudoscalar. 
These processes are often relevant in the same ``tricky'' region of 
parameter space  where  the discrepancies in the predictions of the 
spectrum calculators well exceed the $1\%$ level, Ref.~\cite{Allanach:2003jw}, 
leading to large uncertainties in the relic density prediction.

The influence of these differences on
relic density computations has first been studied in
\cite{Allanach:2004jh} for the Les Houches 2003 workshop.
Since then, all above mentioned programs have undergone major
updates; a re-analysis of the existing uncertainties therefore
seems appropriate.
Moreover, the study of \cite{Allanach:2004jh} concentrated on potentially
large differences along specific lines in the focus point, large $\tan\b$ and
co-annihilation regions. In this article, we consider  the  WMAP allowed
parameter region in the $m_0-m_{1/2}$ plane,
investigating in particular differences in WMAP constraints which arise
from the different SUSY spectrum codes. 
We also address the issue of non-zero $A_0$,
which for very large $A_0<0$ leads to $\st$ co-annihilation.

We first briefly discuss in Section~2 the calculation of the
supersymmetric spectrum.
We then study in Section~3 some specific scenarios:
in Section~3.1 we discuss the case of moderate parameters
(small $m_0$, small to medium $m_{1/2}$, moderate $\tan\b$),
which is most promising for collider phenomenology and where the
calculations are expected to be quite precise. As it turns out there
are, however, non-negligible uncertainties already in this region.
In Section~3.2, we discuss the case of large $\tan\beta$, where
much larger differences are observed.
Section~3.3 then deals with the case of large $m_0$ and 
Section~3.4 with the case of large $m_0$ and large $\tan\beta$.
Here very large uncertainties are found; in particular focus point
behaviour may or may not occur depending on the program. 
The influence of the $A_0$ parameter is discussed in Section~3.5.
In Section~4, we present a web application for online spectrum
comparisons. Finally, Section~5 contains conclusions and an outlook.

For the sake of a fair comparison, we use the same Standard Model (SM) 
input parameters in all programs. In particular, we use
$m_b(m_b)^{\overline{MS}}=4.214$~GeV and $\a_s(M_Z)^{\overline{MS}}=0.1172$
according to \isajet. Moreover, we use a top pole mass of $m_t=175$~GeV
throughout the paper. The parameters of the MSSM are defined following
the SUSY Les Houches Accord (SLHA) \cite{Skands:2003cj}.

We do not discuss here the impact of different cosmological
scenarii. We assume the standard cosmological scenario, in
particular that at the freeze-out temperature when the interaction
rate of particles drops below the expansion rate of the universe,
the universe was radiation dominated. Modifications of the
standard picture for the expansion of the universe could
significantly affect the estimation of the relic density, examples
are models with a low-reheating temperature \cite{Giudice:2000ex}
or with scalar-field kination \cite{salati}.

\section{SUSY spectrum and relic density}

To derive the relic density within a specific SUSY model, mSUGRA for instance,
one needs to compute the mass spectrum and couplings from high-energy input
parameters.
We use the latest version of the four public codes \isajet, \softsusy,
\spheno\ and \suspect\ for this task 
and compare their spectra and the resulting neutralino relic densities. 
These codes basically work as follows:
after specifying the gauge and Yukawa couplings in the $\DRbar$ scheme
at the electroweak scale and starting with an initial guess of the MSSM
parameters, renormalization group (RG) equations are used to run the
parameters to some high scale $M_X$. There boundary conditions are imposed
on the SUSY-breaking parameters, and the couplings and parameters are run
down to the SUSY mass scale. At that scale radiative electroweak symmetry
breaking is checked. The SUSY spectrum is calculated and radiative
corrections are computed. The process is repeated iteratively until a
stable solution is found.
The four programs differ, however, in the implementation of radiative 
corrections (a detailed comparison of the codes can be found in 
\cite{Allanach:2003jw}). 
For one, \isajet, \softsusy
\footnote{Here note that the default option in \softsusy\ 
    is 1-loop running of the squark and slepton mass parameters;
    2-loop running of these parameters has to be switched on by hand. 
    In the following, we always take \softsusy\ with full 2-loop RGE 
    running.}
and \spheno\ apply full 2-loop RG running for all SUSY mass parameters, 
while \suspect\ calculates gaugino and Higgs mass parameters at 2-loops 
but squark and slepton parameters only at 1-loop.
Second, \isajet\ uses step beta functions when passing thresholds 
in the RG evolution, adding additional finite corrections at the end.
In contrast to that the other programs compute the complete 1-loop
threshold corrections at the 
SUSY mass scale $M_{SUSY}=\sqrt{\mst{1}\mst{2}}$.
Third, the use of either on-shell or running masses in the loops
can significantly influence the results even though the
difference is formally a higher-order effect.
Moreover, different approximations are used in some parts of the loop
corrections. 
For example, \isajet\ and \spheno\ apply the complete 1-loop 
corrections given in \cite{Pierce:1996zz} for the neutralino 
and stau masses, while \softsusy\ and \suspect\ use the approximate 
expressions of \cite{Pierce:1996zz} for neutralinos and do not include  
the self-energies for the staus. 
The calculation of the light Higgs mass has recently been standardized
between \softsusy, \spheno\ and \suspect\ to full 1-loop plus leading
2-loop corrections, see \cite{Allanach:2004rh}.
\isajet\ on the other hand uses an 1-loop effective potential, which
typically leads to about 2--3 GeV higher $h^0$ masses compared to
the other programs. Notice, however, that this lies within the present
$\sim$2--3~GeV theoretical uncertainty in $m_h$.
Moreover, as we will see, the exact value of $m_h$ is only important 
in a narrow strip in the large $m_0$ region. 
All considered, we take \isajet, \softsusy\ and \spheno\ as being 
of a comparable level of sophistication as concerns the SUSY 
and heavy Higgs masses.  
Two-loop as opposed to one-loop scalar running, as in \suspect, 
can however have an important influence on the relic density
through differences in the sfermion masses.  

The nature of the LSP, which is a linear combination of the bino
$\tilde B$, wino $\tilde{W}$ and the two higgsino states
$\tilde H_{1,2}$,  is a crucial parameter in the evaluation of the relic
density;
\beqn
  \nt_1= N_{11}\tilde{B}+ N_{12}  \tilde{W} +N_{13}\tilde{H_1}+
  N_{14}\tilde{H_2}
\eeqn
where $N$ is the neutralino mixing matrix.
The LSP-higgsino fraction is given as
\beqn
  f_H=N_{13}^2+N_{14}^2
\eeqn
and is large when the higgsino mass parameter $\mu\lsim M_1,M_2$, 
where $M_1$ and $M_2$ are the U(1) and SU(2) gaugino masses. 
The LSP coupling to the pseudoscalar, $g_{\nt_1\nt_1 A}$, depends 
on the same elements of the neutralino mixing matrix:
\beqn
  g_{\nt_1\nt_1 A} \propto N_{13}^2-N_{14}^2.
\eeqn
The value of $\mu$ depends sensitively,
in certain regions of parameter space, on the SM input
parameters, in particular the top  quark mass and its relation
with the top  Yukawa couplings. 
At large $\m0$, the top Yukawa coupling has a strong influence on 
$m_{H_2}^2$; as a result the $\mu$ parameter becomes very sensitive 
to $h_t=\sqrt{2} \hat m_t/v_2$, where $\hat m_t$ is the running t-quark 
mass and $v_2$ is the vev of the second Higgs doublet:
\begin{equation}
  \mu^2 = \frac{(\overline m_{H_1}^2-\overline m_{H_2}^2\tan^2\beta)}
               {\tan^2\beta-1} - \frac{1}{2}M_Z^2 \,.
\end{equation}
Here $\overline m_{H_i}^2= m_{H_i}^2-t_i/v_i$, $i=1,2$,
with $t_i$ the tadpole contributions. 
See \cite{Allanach:2003jw} for more detail. For the intermediate
to large values of $\tan\beta$ that we will consider, the
$m_{H_2}^2$ term dominates in the extraction of $\mu$. Differences
in the $\mu$ parameter will affect the neutralino couplings, in
particular the coupling to the pseudoscalar $A$. In the mass spectrum 
these differences most obviously show up as differences in the mass 
of the neutralino that is dominantly higgsino, usually $\nt_3$.
The programs under consideration all apply the 1-loop 
corrections of \cite{Pierce:1996zz} plus the 2-loop QCD corrections
of \cite{Avdeev:1997sz}. Nevertheless the differences in $h_t$ are large enough 
to lead to huge discrepancies in $\mu$ at large $m_0$.

The mass of the pseudoscalar, $\ma$, is another important parameter in the
computation of the relic density. This mass also depends  sensitively  on the
SM input parameters, in particular the  bottom quark
mass and its translation to the bottom Yukawa coupling. The bottom Yukawa coupling
which is large at high $\tan\beta$ impacts the Higgs sector since
$m_{H_1}^2$ is driven by $h_b=\sqrt{2}\hat m_b/v_1$, where $\hat
m_b$ is the running b-quark mass and $v_1$ is the vev of the first
Higgs doublet. The physical pseudoscalar mass directly depends on
$m_{H_1}^2$:
\begin{equation}
   \ma^2 = \frac{1}{\cos{2\beta}}
           (\overline m_{H_1}^2- \overline m_{H_2}^2)
           + \frac{s^2_\beta t_1}{v_1} + \frac{c^2_\beta t_2}{v_2}-M_Z^2 \,.
\end{equation} 
The four spectrum codes all apply the corrections of 
\cite{Pierce:1996zz,Avdeev:1997sz}, resumming the 1-loop SUSY corrections 
according to \cite{Carena:1999py}. This brings in general good agreement 
on $\ma$; however as we will see the remaining differences can still lead 
to sizable discrepancies in $\Omega$ in parts of the parameter space.

\section{Results}

\subsection{\mbf Small $m_0$, small to medium $m_{1/2}$, moderate $\tan\b$}

We start out with an easy, collider-friendly scenario of
small $m_0$, small to medium $m_{1/2}$ and moderate $\tan\b$.
Such a scenario has gluinos and squarks with masses up to 1~TeV
which cascade-decay into neutralinos and sleptons. It can hence
provide the favourite LHC signature of jets plus same-flavor
opposite-sign leptons. It also has gauginos and sleptons within
the kinematical reach of a future $e^+e^-$ linear collider (ILC)
and is thus very well suited for both LHC and ILC studies.

\begin{figure}[t]
\centerline{\epsfig{file=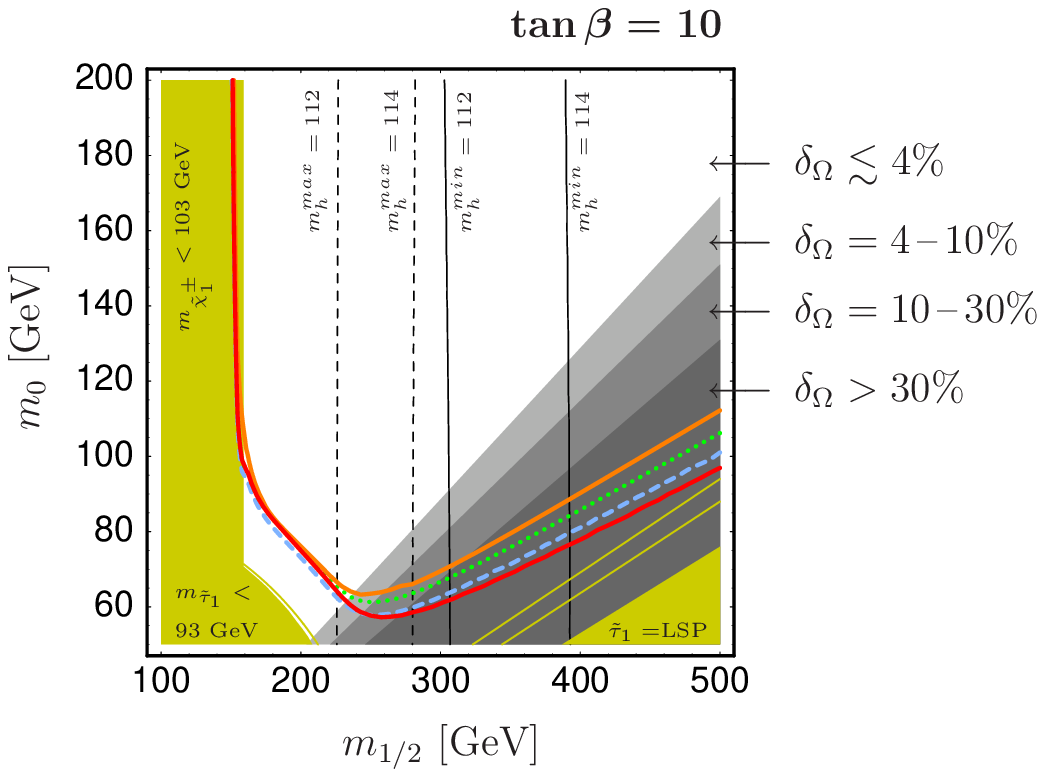, height=8cm}}
\caption{Comparison of results in the $m_0$--$m_{1/2}$ plane, for
$A_0 = 0$, $\tan\beta=10$, $\mu>0$, and $m_t=175$\,GeV. The red (dark) 
and orange (light) full lines show the variation of the $2\s$ upper limit
$\Omega<0.1287$ when \micromegas\ is linked to 
\isajet, \softsusy\ or \spheno. 
The orange line basically comes from \softsusy\ while the red 
one comes from \isajet. In addition, the upper bound from \spheno\ 
is shown as green dotted line, and that of \suspect\ as blue dashed line.
The light, medium and dark gray shaded areas show the regions where
the relative differences in $\Omega$, $\delta\Omega$ of eq.~\eq{domega}, 
are 4--10\%, 10--30\% and $>$30\%, respectively. 
Also shown are contours of minimal
(full black lines) and maximal (dashed black lines) $h^0$ masses as
obtained by the spectrum codes. The yellow region on the left
is excluded by LEP2 constraints; in the yellow triangle in the
bottom right corner $m_{\ti\tau_1}<\mnt{1}$ in \isajet. 
The yellow lines show the boundaries of the excluded region in the 
other codes.}
\label{fig:tb10}
\end{figure}

In the region considered in this section, as in most of the mSUGRA
parameter space, the LSP is nearly a pure bino.
As such it couples preferably to
right-chiral sfermions with a coupling proportional to the hypercharge.
The main annihilation channel for the LSP is then into lepton pairs
via t-channel exchange of right-chiral sleptons.
This process is  efficient enough to meet the WMAP upper limit only
in the low $\m0$--$\mhf$ corner of the parameter space, the so-called
bulk region.  Indeed, for a pure bino LSP the relic density is
approximately $\Omega \propto \mslr^4/\mneuto^2$, implying that both
the $\tilde{l}_R$ and the $\nt_1$ must be light. Since sleptons must be
beyond the reach of LEP2, the upper limit from WMAP is only satisfied
in a very small region below $\mhf\sim 240$~GeV, see \fig{tb10}.
The bulk region is, however, associated with a light Higgs below the LEP2
limit {\footnote {Note that an increase in the top-quark mass loosens the
LEP2 constraint from the light Higgs.}}.
Light neutralinos can also annihilate efficiently into fermion pairs
near a $Z$ or Higgs resonance. This corresponds to the near vertical
WMAP line in \fig{tb10}. This possibility is however by large excluded
by the LEP direct limits \cite{lepsusywg} on chargino pairs, 
which in effect translate
into a lower limit on the LSP mass in mSUGRA.

Agreement with WMAP is recovered for heavier neutralinos
($\mhf\gsim 240$ GeV) with the additional contributions from
co-annihilation channels, the so-called co-annihilation region.
For co-annihilation to be effective, the mass difference between
the slepton NLSP and the $\nt_1$ LSP  must be rather small (less than
$\sim$10 GeV). Such degenerate sleptons/neu\-tra\-linos are found in
the low $\m0$ region of mSUGRA. The $\stauo$ is the lightest
slepton due both to the effect of the $\tau$-Yukawa coupling in
the RGE running of $m_{\stau_R}$ as well as to the mixing between
$\tilde{\tau}_L$ and $\tilde{\tau}_R$ which lowers the mass of the
$\stauo$ to $\mstau{1} <m_{\stau_R}$.
In fact, co-annihilation processes with $\stauo$
dominate over most of the allowed region in \fig{tb10}.
In the co-annihilation region, it was shown in Ref.~\cite{Allanach:2004xn}
that the relic density is extremely sensitive to the mass difference
between $\stauo$ and $\nt_1$, $\Delta M(\neuto\stauo)$.
Typically a shift in $\Delta M(\neuto\stauo)\approx 1$~GeV induces
$\Delta\Omega\approx 10\%$.
Previous comparisons between the public SUSY spectrum 
codes~\cite{Allanach:2003jw} have shown that the predicted masses 
often differ by more than $\pm 1\%$, inducing discrepancies in 
$\Delta M(\neuto\stauo)$ above $\pm 1$ GeV and
hence large uncertainties in the relic density.

%

These expectations are corroborated by 
a scan in the  $m_0$--$m_{1/2}$ plane comparing the predictions
of the four spectrum codes.
\Fig{tb10} shows results for $A_0 = 0$, $\tan\beta=10$, $\mu>0$ and
$m_t=175$\,GeV.
The red and orange lines show the variation of the $2\s$ upper
limit $\Omega<0.1287$ when \micromegas\ is linked to
\isajet, \softsusy\ and \spheno.
In addition, the light, medium and dark gray shaded areas show the regions
where the relative differences in $\Omega$,
\begin{equation}
  \d_\Omega\equiv (\Omega_{max}-\Omega_{min})/\Omega_{mean},
\label{eq:domega}
\end{equation}
are 4--10\%, 10--30\% and $>$30\%, respectively. 
Here $\Omega_{max}$ and $\Omega_{min}$ are the maximal and minimal
values and $\Omega_{mean}$ the arithmetic mean of the $\Omega$
values obtained from \isajet, \softsusy\ and \spheno\ at a specific 
parameter point. 
We do not include \suspect\ in the calculation of $\d_\Omega$ because
it has only 1-loop scalar running. 
A $\d_\Omega$ of 30\% corresponds to the present precision of WMAP, 
while the PLANCK experiment \cite{PLANCK} is expected to reach a precision of 4\%, 
corresponding to the white area in \fig{tb10}.
Also shown are contours of the minimal and maximal
$h^0$ masses as obtained by the four spectrum codes.
As a general rule, the $m_h^{max}$ lines come from \isajet, while
the $m_h^{min}$ lines come from the other programs.
Note that the bulk region is practically excluded.

The red (maximal $\Omega$) and orange (minimal $\Omega$) lines 
in \fig{tb10} come from \isajet\ and \softsusy, respectively. 
The values obtained from \spheno, shown as dotted green line, lie 
in between these curves. 
In the co-annihilation region  the results of \suspect, 
shown as dashed blue line, fall within the red and orange lines. 
While the differences in the WMAP bounds in \fig{tb10} do 
not look dramatic, it becomes clear from the grey shaded areas that
the relative differences in $\Omega$ are quite large in the
allowed parameter space, that is in the co-annihilation region, 
where the precise mass differences, in particular between $\ti\tau_1$ 
and $\nt_1$, are important. 
 
\isajet, \softsusy\ and \spheno\ typically agree on the $\stauo$ mass to 
$\sim$1\%. The difference mainly comes from \softsusy, which neglects 
the tau Yukawa coupling $h_\tau$ and the $\stau$ self-energy correction 
and hence gets a 
slightly smaller $m_{\stau_1}$. \softsusy, \spheno~ and \suspect\ 
agree very well on the $\neuto$ mass, while \isajet\ finds a 
$\mnt{1}$ smaller by about 2\%.
As a consequence, both \softsusy\ and \spheno\ tend to give a smaller 
$\stau_1$--$\nt_1$ mass difference than the other two programs,
and hence a  smaller $\Omega$ in the co-annihilation region.
As an example, \Tab{masses1} lists the relevant masses together with $\Omega$
for $m_0=70$~GeV, $m_{1/2}=350$~GeV, $A_0=0$, $\tan\b=10$ and $\mu>0$.
\Tab{relcontribs} gives the according relative contributions to $\Omega^{-1}$
for this point. Note here the contribution of the  
$\stau_1$ and $\ser,\smur$ co-annihilation channels.
Clearly our expectations that the mass difference is the most
important parameter are confirmed. The 2~GeV decrease in $\Delta M(\neuto\stauo)$
when going from \spheno\ to \softsusy\ roughly corresponds to a decrease 
of ${\cal O}(20\%)$ in $\Omega$ as expected. 
As a result of the mass spectrum,
one finds a larger contribution from the co-annihilation channels
for \softsusy\ where it amounts to almost 80\% of the effective annihilation
cross section as compared to the other codes where
co-annihilation channels contribute $\sim$50--70\%.
\isajet, which agrees well with \spheno\ on the $\stauo$ mass but finds  
a smaller $\mnt{1}$, has the largest $\neuto$--$\stauo$ mass difference 
and a $\sim$50\% higher $\Omega$ as compared to \spheno. 
For similar $\Delta M(\neuto\stauo)$, \isajet\ predicts a slightly lower 
value for the relic density as compare to other codes because of a lower LSP mass.
\suspect\ on the other hand agrees well with \softsusy/\spheno\ on the 
LSP mass, but due to the missing 2-loop effects in the running of the 
slepton masses it gets a heavier $\stauo$ (and $\ti e_R$) and hence a 
larger $\Omega$. 
We have checked that when using only 1-loop RGEs for the slepton 
mass parameters in \softsusy, it reproduces the results of \suspect.

%
\begin{table}[t]\begin{center}\vspace*{4mm}
\begin{tabular}{c|cccc}
                & \isajet\ & \softsusy\ & \spheno\ & \suspect\ \\
\hline
   $\nt_1$    & 136.7 & 140.0 & 139.5 & 140.0 \\
   $\stau_1$  & 147.7 & 145.7 & 147.1 & 149.7 \\
   $\ti e_R$  & 155.7 & 153.8 & 155.4 & 157.6 \\
   $h^0$      & 115.8 & 113.1 & 113.4 & 113.3 \\
\hline
   $m_{\stau_1}-\mnt{1}$ & 11.0 & 5.7 & 7.6 & 9.7 \\
\hline
   $\Omega$ & 0.136 & 0.069 & 0.092 & 0.120 \\
\hline
\end{tabular}\end{center}
\caption{Relevant masses, the $\nt_1$--$\stau_1$ mass
         difference (in GeV) and the resulting $\Omega$ for $m_0=70$~GeV,
         $m_{1/2}=350$~GeV, $A_0=0$, $\tan\b=10$ and $\mu>0$.
         The higgsino fraction of $\nt_1$ is 1.4--1.5\% in all cases.}
\label{tab:masses1}
\end{table}

%
\begin{table}[t]\begin{center}
\begin{tabular}{l|cccc}
   channel  & \isajet\ & \softsusy\ & \spheno\ & \suspect\ \\
\hline
   $\nt_1\nt_1\to ee$            & 28\% & 10\% & 16\% & 22\% \\
   $\nt_1\nt_1\to \tau\tau$      & 16\% &  6\% &  9\% & 13\% \\
   $\nt_1\ti e_R\to \g/Z\, e$    &  8\% &  8\% &  8\% & 10\% \\
   $\nt_1\stau_1\to \g/Z\,\tau$  & 30\% & 39\% & 38\% & 34\% \\
   $\stau_1\stau_1\to \tau\tau$  &  5\% & 17\% & 11\% &  7\% \\
   $\stau_1\stau_1\to \g\g,\g Z$ &  2\% &  7\% &  6\% &  3\% \\
   $\ti e_R\stau_1\to e\tau$     &  2\% &  6\% &  4\% &  2\% \\
\hline
\end{tabular}\end{center}
\caption{Relative contributions to $\Omega^{-1}$ for $m_0=70$~GeV,
$m_{1/2}=350$~GeV, $A_0=0$, $\tan\b=10$ and $\mu>0$;
with $e\equiv e,\mu$ and $\ti e_R\equiv\ti e_R,\ti \mu_R$.}
\label{tab:relcontribs}
\end{table}

Some more comments are in order.
First, a non-zero value of $A_0$ shifts the contours of constant
Higgs masses and moves the position of the stau co-annihilation
strips as well as of the excluded regions in \fig{tb10}; it does however
not change the picture qualitatively,
provided $A_0$ is not so large as to make $\st_1$ the (N)NLSP.
The case of a non-zero $A_0$ will be discussed in detail in Section~3.5. 
Second, for the reference point SPS1a' of the SPA project \cite{SPA},
($m_0=70$~GeV, $m_{1/2}=250$~GeV, $A_0=-300$, $\tan\b=10$, $\mu>0$
and $m_t=178$~GeV), \isajet, \softsusy\ and \spheno\ 
give values of $\Omega=0.126$, 0.103 and 0.114, respectively.
\softsusy\ with 1-loop scalar running  
gives $\Omega=0.125$, and \suspect\ $\Omega=0.126$. 
All values lie within the WMAP allowed range of Eq.~\eq{wmap}
at this point, with the spread of $\d\Omega\simeq20\%$ again being
mainly due to $\Delta M(\neuto\stauo)$.

\subsection{\mbf Large $\tan\b$}

We next consider large values for $\tan\beta$; we stay however within
collider-friendly scenarios with small $\m0$ and small to medium $\mhf$.
At large values of $\tan\beta$, the enhanced couplings of the heavy
Higgses to $b\bar{b}$ and $\tau\tau$ lead to an enhancement of neutralino
annihilation channels through $\nt_1\nt_1\to H^0\!,A^0\to b\bar b,\tau\tau$.
Because of the Majorana nature of the LSP the main contribution is
the pseudoscalar exchange, the CP-even state being P-wave suppressed.
Even though the LSP is mostly bino, its small higgsino component
is sufficient to make  annihilation through the pseudoscalar and the
Goldstone component of $Z$ exchange dominant.
These contributions are added to the contributions from t-channel
sfermion exchange or from  co-annihilation with staus that were
already present at lower values of $\tan\beta$.

\begin{figure}[p]
\centerline{\epsfig{file=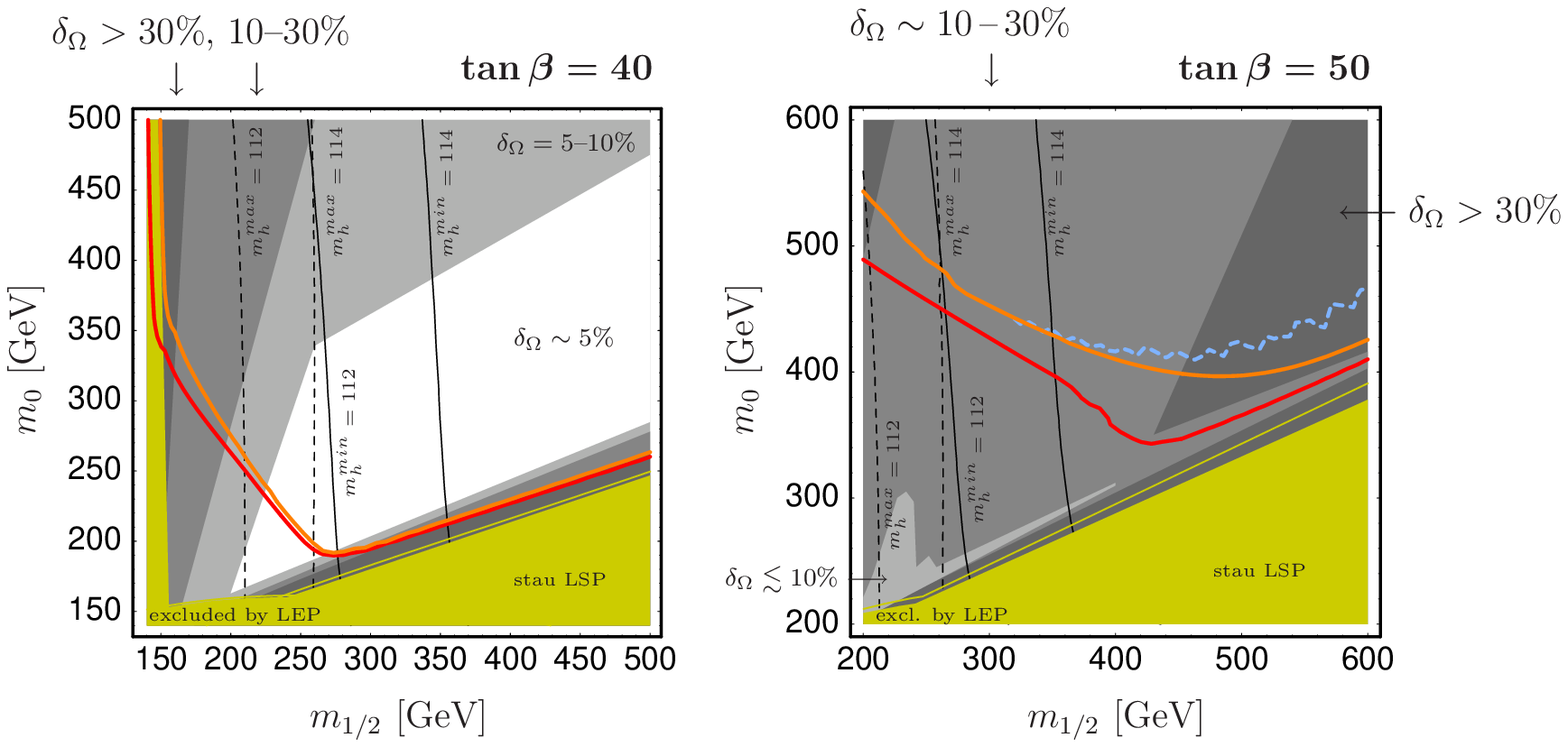, height=8cm}}
\caption{Comparison of results analogous to \fig{tb10}
         but for $\tan\beta=40$ (left) and $\tan\beta=50$ (right);
         $A_0 = 0$, $\mu>0$, and $m_t=175$\,GeV.
         The red and orange lines show again the variation of the bound
         $\Omega<0.1287$ due to differences in the spectra from 
         \isajet, \softsusy\ and \spheno. 
         In the right panel, the blue dashed line shows in addition 
         how the upper curve would move when including \suspect.}
\label{fig:tb4050}
\end{figure}

At low $m_{1/2}$, the  annihilation into $b\bar{b}$  typically
constitutes more than 80\% of the effective annihilation cross
section. For a fixed value of $\mhf$, hence of neutralino mass,
the relic density decreases with $\m0$ since both the sfermion
masses as well as the pseudoscalar mass decrease, making for more
efficient annihilation. Because of the enhanced contribution of
the pseudoscalar exchange, a much larger region of parameter space
in the bulk is compatible with the WMAP upper bound as compared to
intermediate $\tan\beta$ values, see \fig{tb4050}. Nevertheless as
one moves towards larger values of $\mhf$ and a heavier LSP, one
must again appeal to co-annihilation to retain consistency with
WMAP, leading to a mixed region with both co-annihilation and 
pseudoscalar exchange. The co-annihilation occurs exclusively with
$\stauo$, which is much lighter than the other sleptons at large
$\tan\b$. Note that for the range of $\mhf$ which we are
considering, we are never near the heavy Higgs pole.

The relic density is again sensitive to $\Delta M(\neuto\stauo)$
for the co-annihilation processes.
Sensitivity to $\Delta M(\neuto A) = m_A-2\mnt{1}$ as well as to
the $\neuto\neuto A$ coupling are expected for the Higgs contribution,
see \cite{Allanach:2004xn}.
As already mentioned, the bottom and tau Yukawa couplings play an important
role in radiative corrections to the sparticle and Higgs masses at
large $\tan\beta$, leading to larger differences in the spectra.
Consequently in the computation of the relic density we also observe
larger discrepancies between the four codes.

%

%
\begin{figure}[p]
\centerline{\epsfig{file=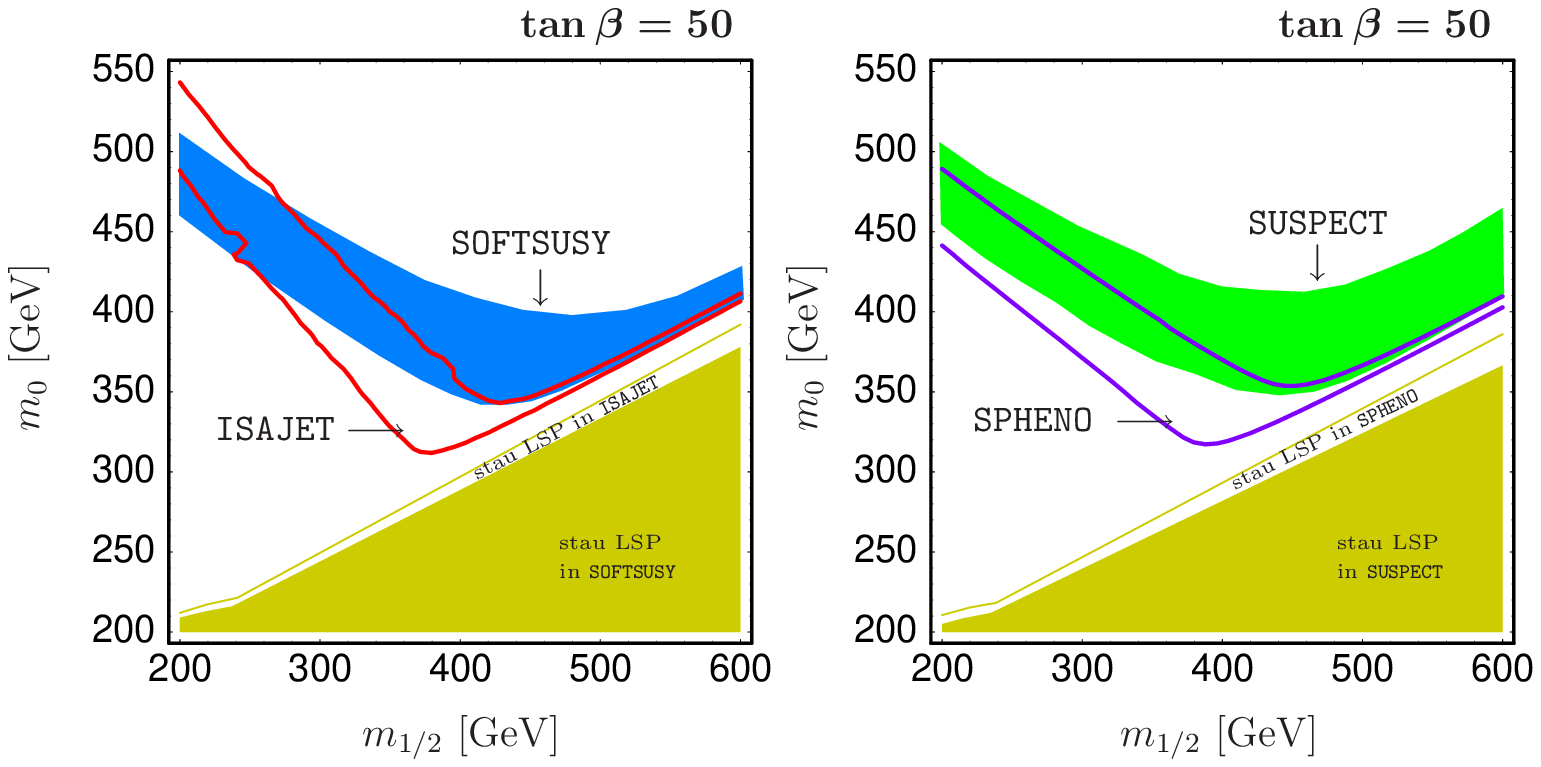, height=8cm}}
\caption{WMAP allowed regions of $0.0945\leq\Omega\leq 0.1287$ 
         for $\tan\beta=50$, $A_0 = 0$, $\mu>0$, $m_t=175$\,GeV;
         left: \isajet\ and \softsusy,
         right: \spheno\ and \suspect.}
\label{fig:tb50regions}
\end{figure}

\Fig{tb4050} compares the results of the various codes in the
$m_0$--$m_{1/2}$ plane analogous to \fig{tb10} but for
$\tan\beta=40$ (left) and $\tan\beta=50$ (right).
The other parameters are $A_0=0$, $\mu>0$ and $m_t=175$~GeV as before.
At $\tan\beta=40$, the WMAP exclusion curves seem to agree quite well.
Small differences ($\Delta \Omega<5\%$) are observed over much of
the plane, but these increase rapidly to 10--30\% and more as one
moves into the WMAP allowed region.
Near the stau-LSP border, differences in the predictions of the
spectrum calculators for the $\stau_1$ masses, and hence for
$m_{\stau_1}-\mnt{1}$, explain this discrepancy, just as was the
case for $\tan\b=10$. Large differences are also observed for low
$\mhf$ in the region near the band excluded by LEP limits. These
discrepancies are due to differences in $m_{\nt_1}$.  Specifically
some codes allow a significant annihilation rate through the light
Higgs exchange in a region that is allowed by the LEP limit on
charginos. 
Here again the low $\mhf$ (bulk) region is not compatible 
with the lower limit on the Higgs mass.

For further illustration, we pick a parameter point from
\fig{tb4050}a, $(m_0,\,m_{1/2})=(194,\,300)$ GeV at
$\tan\beta=40$. Details on the spectrum relevant for the relic
density calculation and the list of important channels for all
four codes are presented in Table 3. For this parameter choice we
are in a mixed region where both co-annihilation and Higgs
exchange processses are important. All codes agree quite well on
the values of $m_{\nt_1}$, $\ma$ and consequently $\ma - 2m_{\nt_1}$
with maximal variation on the latter of about 5\%. The 
variation in the  $\mu$ parameter is below 3\%. 
The variation of the NLSP--LSP mass difference is 2\% within \isajet, 
\softsusy\ and \spheno, but 40\% if we also include \suspect.
The difference in the $\stauo$ mass between \softsusy\ and \spheno\ 
can again be explained by the missing $h_\tau$ and $\stau$ self-energy    
corrections in the former program, which is roughly a 1\% effect.
It is interesting to note that this also influences $\ma$ at the level 
of few per-mille. 
All considered, the uncertainties in the Higgs annihilation and the 
stau co-annhilation channels are of similar importance in \isajet, 
\softsusy\ and \spheno, leading to a spread in $\Omega$ of 25\% for 
the parameter point of Table~3. 
If we interpret this as $\Omega=0.107\pm 0.013$, then \suspect\ 
deviates by $2.7\sigma$ due to its larger NLSP--LSP mass difference.

%
\begin{table}[p]\begin{center}
\begin{tabular}{l||cccc}
     & \isajet\ & \softsusy\ & \spheno\ & \suspect\ \\
\hline
\hline
   \qquad $\nt_1$    & 117.2 & 119.9 & 119.7 & 119.9 \\
   \qquad $\stau_1$  & 131.4 & 133.2 & 131.4 & 137.7 \\
   \qquad $h^0$      & 115.3 & 112.7 & 113.0 & 112.8 \\
   \qquad $A^0$      & 363.4 & 363.2 & 366.4 & 364.4 \\
   \qquad $\nt_3$    & 394.9 & 401.4 & 405.3 & 405.3 \\
\hline
   $m_{\stau_1}-\mnt{1}$ & 14.2 & 13.3 & 11.6 & 17.8 \\
   $m_A-2\mnt{1}$        & 129 & 123 & 127 & 125 \\
\hline
   $\nt_1\nt_1\to b\bar b$           & 40\% & 38\% & 30\% & 49\% \\
   $\nt_1\nt_1\to ee      $          & 12\% & 10\% & 10\% & 14\% \\
   $\nt_1\nt_1\to \tau\tau$          & 17\% & 14\% & 13\% & 19\% \\
   $\nt_1\stau_1\to h\tau$           & 13\% & 16\% & 21\% &  7\% \\
   $\nt_1\stau_1\to \gamma/Z\,\tau$  & 12\% & 14\% & 18\% &  7\% \\
   $\stau_1\stau_1\to hh$            &  1\% &  2\% &  3\% &  -- \\
\hline
   \qquad $\Omega$ & 0.120 & 0.107 & 0.094 & 0.142 \\
\hline
\end{tabular}
\end{center}
\caption{Masses and mass differences (in GeV),
   the most important contributions, and the resulting $\Omega$
   for $m_0=194$~GeV, $m_{1/2}=300$~GeV, $A_0=0$, $\mu>0$ and $\tan\b=40$.
   The higgsino fraction of $\nt_1$ is 1.8\% in all cases.}
\label{tab:tanbe40}
\end{table}

%
\begin{table}[p]\begin{center}
\begin{tabular}{l||cccc}
      & \isajet\ & \softsusy\ & \spheno\ & \suspect\ \\
\hline
\hline
   \qquad $\nt_1$    & 139.1 & 142.2 & 141.9 & 142.1 \\
   \qquad $\stau_1$  & 208.7 & 217.6 & 214.6 & 223.3 \\
   \qquad $h^0$      & 116.3 & 113.9 & 114.3 & 114.1 \\
   \qquad $A^0$      & 369.0 & 366.2 & 371.9 & 365.3 \\
   \qquad $\nt_3$    & 449.9 & 457.3 & 462.7 & 463.1 \\
\hline
   $m_{\stau_1}-\mnt{1}$ & 70 & 75 & 73 & 81 \\
   $m_A-2\mnt{1}$        & 91 & 82 & 88 & 81 \\
\hline
   $\nt_1\nt_1\to b\bar b$           & 81\% & 83\% & 82\% & 83\% \\
   $\nt_1\nt_1\to \tau\tau$          & 15\% & 14\% & 14\% & 14\% \\
\hline
   \qquad $\Omega$ & 0.104 & 0.087 & 0.102 & 0.088 \\
\hline
\end{tabular}
\end{center}
\caption{Same as \tab{tanbe40} but for $m_0=m_{1/2}=350$~GeV and $\tan\b=50$.
         The higgsino fraction of $\nt_1$ is 1.4\%.}
\label{tab:tanbe50}
\end{table}

Increasing further $\tan\b$ means that the $Ab\bar{b}$ coupling is
further enhanced and pseudoscalar exchange dominates over most of
the probed parameter space. \Fig{tb4050}b compares the various
codes for $\tan\b=50$. In this case, large discrepancies are found
in the relic density and this over most of the parameter space.
In particular, the minimal and maximal upper boundaries from WMAP
displayed as red and orange curves differ significantly. As a result
of the pseudoscalar exchange contribution, the bulk region is much 
larger compared to small $\tan\beta$. 
We however stress that this bulk region is not one of
t-channel sfermion exchange but rather one of heavy Higgs
annihilation. Stau co-annihilation also plays a role near the stau-NLSP
boundary; however this typically drives the relic
density below the WMAP range. Large $\tan\beta$ also means
larger discrepancies between the predictions of the spectrum
calculators especially for the pseudoscalar mass.
The large discrepancies in $\Omega$ over most of
the parameter space are due to the differences in $\ma-2\mnt{1}$.
Typically, at such large $\tan\beta$ the mass of the pseudoscalar in 
\softsusy\ and \suspect\ is lighter than that predicted by the other two codes.
The only region where $\delta \Omega<10\%$ lies  at small $\mhf$  where the 
pseudoscalar exchange
diagram is less important and better agreement for the masses is found. 

\Fig{tb50regions} explicitly compares the 2$\sigma$ WMAP allowed regions
of the four programs at $\tan\beta=50$. The difference in the prediction
of the pseudoscalar mass also explains why the band near the stau
co-annihilation region is much narrower for \isajet\ and \spheno.
Here one is sitting too far from the Higgs resonance to get a significant
contribution to the annihilation cross-section, the only remaining WMAP
allowed region being the narrow stau co-annihilation strip.
For further illustration, we pick a parameter point from
\fig{tb50regions}, $(m_0,\,m_{1/2})=(350,\,350)$ GeV at $\tan\beta=50$. The
results for this point are listed in Table~4.
Here the mass difference between the NLSP and the LSP is much too
large to get a significant contribution from co-annihilation processes.
The main channels are annihilation of neutralinos into fermion pairs 
via pseudoscalar exchange. 
Although one is far from the Higgs resonance, this 
process is efficient enough due to the enhanced coupling of the 
Higggs to $b\bar{b}$ and $\tau\tau$.
For this point one gets as usual rather good agreement among all
codes in the $\nt_1$ masses and in the higgsino fraction.  The
pseudoscalar masses also agree within 1--2\%; for the
resonance parameter $\ma-2m_{\nt_1}$ the differences are however 
around 10\%.  
In \cite{Allanach:2004xn} it was shown that in this region
a 4\% shift in $\ma-2m_{\nt_1}$ leads to a 10\% change in
$\Omega$. The discrepancies in the mass difference found in
Table~4 explain the difference between the value of the relic density in
\spheno, \softsusy\ and \suspect. In the case of \isajet\ the
decrease in the annihilation cross section  due to the fact that
one is sitting further away from the Higgs resonance is partly
compensated by a lower value of the $\mu$ parameter (to wit the
smaller value of $m_{\nt_3}$) hence a larger $\nt_1\nt_1 A$
coupling. 
We have also checked explicitly that by adjusting $\ma-2 m_{\nt_1}$ 
to the \spheno\ value in \softsusy\  we recover very good agreement
between the two programs.

To put these results in perspective, we also remark that there is
a strong $m_b(m_b)$ dependence in the computation of the
pseudoscalar Higgs mass as discussed in Section~2.
This has an impact on the relic density \cite{Gomez:2004ek,Allanach:2004xn}.
For  example for the parameters of Table \ref{tab:tanbe50},
decreasing $\mbmb$ to 4.168~GeV (less than a 2\% change) makes the
result of \softsusy\ agree perfectly with the ones from \isajet.
Considering that there are large theoretical uncertainties in the
extraction of $\mbmb$, this source of uncertainty at present
exceeds the one estimated by taking the difference between codes.

\subsection{\mbf Large $m_0$, focus point}

\begin{figure}[p]
\centerline{\epsfig{file=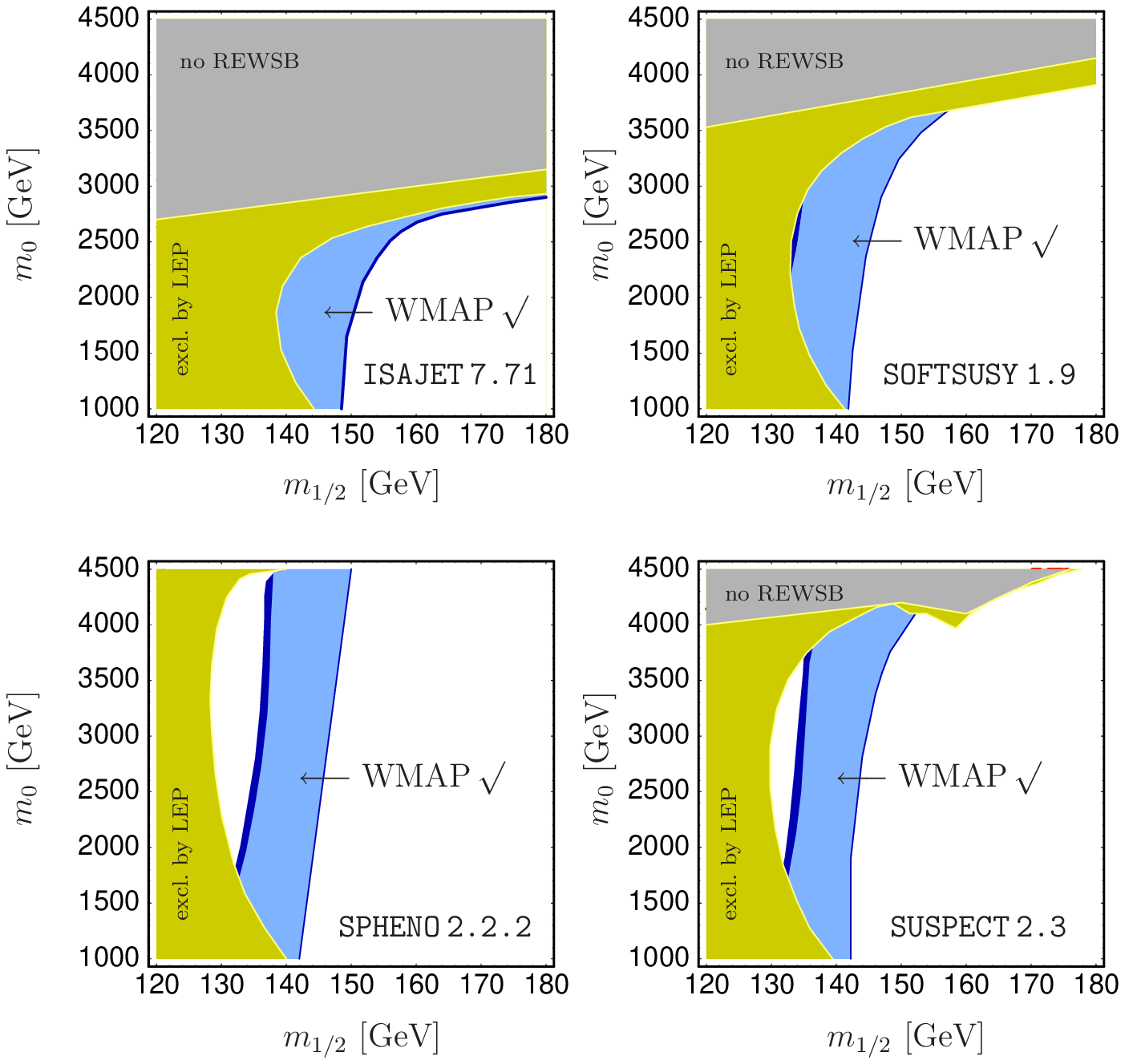, height=14cm}}
\caption{WMAP allowed regions (blue) in the $m_0$--$m_{1/2}$ plane for
         large $m_0$; $\tan\beta=10$, $A_0 = 0$, $\mu>0$, $m_t=175$\,GeV.
         In the dark blue bands $0.0945 \leq \Omega \leq 0.1287$,
         while in the light blue bands $\Omega < 0.0945$.
         In the gray areas there is no radiative EWSB; the yellow regions
         are excluded by the LEP bound $\mch{1}>103$~GeV.}
\label{fig:tb10focus}
\end{figure}

Large $m_0$ is a notoriously difficult region which suffers from
large uncertainties. The reason is the extreme sensitivity of the
$\mu$ parameter to the top Yukawa coupling alluded to in Section~2. 
We limit our discussion to gaugino and higgsino masses within
the reach of LHC and ILC and consider values of $m_0$ up to 4.5~TeV.
\Fig{tb10focus} shows the allowed
regions in the $m_0$--$m_{1/2}$ plane for $m_0=1$--4.5~TeV,
$\tan\b=10$, $A_0 = 0$, $\mu>0$ and $m_t=175$\,GeV. A striking
discrepancy between the codes is the occurrence or non-occurrence
of focus-point behaviour and related with this the limit of
radiative electroweak symmetry breaking (REWSB). In
\fig{tb10focus}, the four programs agree more or less up to
$m_0\sim 2$~TeV. Above this value, the results of \isajet\ become
very different, 
with REWSB breaking down around $m_0\sim 2.7$--3~TeV.
In \softsusy\ and in \suspect\ this happens only around $m_0\sim 3.5$--4~TeV
while in \spheno\ one can go to much higher $m_0$.
In fact this  behaviour is related to small differences in the
treatment of the top Yukawa coupling; focus point behaviour can be
recovered for all codes when one lowers the top-quark mass.

In the allowed parameter space, the main annihilation channel for neutralinos
is into fermion pairs. Consistency  with WMAP then requires some enhancement
factor for the annihilation cross section. This is in principle provided by
the light Higgs resonance ---but only in a narrow strip of the parameter space.
The relic density hence becomes very sensitive to the $m_h-2\mnt{1}$
mass difference (as compared to the decoupling temperature of the 
neutralinos, $T_f\approx \mnt{1}/25$). The width of the $h^0$ is not an 
important parameter because it is much smaller then $T_f$. 

When the $\nt_1$ mass is slightly below half the $h^0$ mass, most of
the $\nt_1$'s annihilate efficiently through $\nt_1\nt_1\to h^0\to b\bar b$.
This requires a very small $m_{1/2}$, roughly $m_{1/2}\lsim 150$~GeV as
can be seen in \fig{tb10focus}. On the other hand, the LEP bound of
$\mch{1}>103$~GeV requires $m_{1/2}\gsim 130$--140~GeV.
In \fig{tb10focus} the bands that are within the $2\sigma$ WMAP range correspond 
to $m_h-2\mnt{1}$ either of a few hundred MeV or around 10~GeV
(15~GeV in case of \isajet). In between these values, the Higgs annihilation
mechanism is too efficient, resulting in $\Omega < 0.0945$.
\Tab{largem0} gives examples for $m_0=2$~TeV and $m_0=3.8$~TeV.
For $m_0=2$~TeV, \softsusy, \spheno\ and \suspect\ predict similar masses
and LSP higgsino fractions. As expected, the relic density decreases
as one moves slightly away from the pole.
For \isajet, predictions for the relic density for a given $m_h-2\mnt{1}$
are typically lower than for the other codes, since two other
effects enhance the annihilation cross-section: a larger LSP higgsino fraction
and the fact that with a lighter LSP one benefits from the Z-exchange
contribution.
For $m_0=3.8$~TeV, only \softsusy, \spheno\ and \suspect\ find viable
RGE solutions. There are now large $\sim 60\%$ discrepancies in the
$\mu$ parameter also among these codes. This is reflected in quite 
different higgsino fractions, and in turn in ${\cal O}(100\%)$
differences in the values of $\Omega$.
As a side remark we note that the large uncertainties in the $\mu$
parameter also lead to significant discrepancies in the $\nt_{3,4}$ and
$\ch_{1,2}$ masses, which can  considerably impact the collider 
phenomenology of a particular mSUGRA point.

%

%
\begin{table}[t]\begin{center}
{\bf\mbf $m_0=2$~TeV}\\[2mm]
\begin{tabular}{l||cccc}
                & \isajet\ & \softsusy\ & \spheno\ & \suspect\ \\
\hline
\hline
   \quad $\nt_1$   &  54.9 &  57.8 &  58.2 &  58.2  \\
   \quad $h^0$     & 115.9 & 116.5 & 116.9 & 116.7  \\
   \quad $\nt_3$   & 290.5 & 383.9 & 450.6 & 441.5  \\
\hline
   $m_h-2\mnt{1}$       & 6.1 & 0.9 & 0.5 & 0.3 \\
   $f_H(\nt_1)$         & 4.0\% & 1.9\% & 1.3\% & 1.3\% \\
\hline
   $\nt_1\nt_1\to b\bar b$     &  90\% &  90\% & 90\% & 90\% \\
   $\nt_1\nt_1\to \tau\tau$    &  9\% &  9\% & 9\% & 9\% \\
\hline
   \quad $\Omega$ & 0.011 & 0.011 & 0.023 & 0.038 \\
\hline
\end{tabular}\\[6mm]
{\bf\mbf $m_0=3.8$~TeV}\\[2mm]
\begin{tabular}{l||cccc}
                  & \isajet\ & \softsusy\ & \spheno\ & \suspect\ \\
\hline
\hline
   \quad $\nt_1$   & -- &  56.1 &   59.2 &  57.8  \\
   \quad $h^0$     & -- & 125.8 &  122.1 & 121.6  \\
   \quad $\nt_3$   & -- & 243.7 &  450.7 & 301.9  \\
\hline
   $m_h-2\mnt{1}$       & -- & 13.6 & 3.7 & 6.0 \\
   $f_H(\nt_1)$         & -- & 6.2\% & 1.3\% & 3.4\% \\
\hline
   $\nt_1\nt_1\to b\bar b$     &  -- &  82\% & 90\% & 90\% \\
   $\nt_1\nt_1\to \tau\tau$    &  -- &   9\% &  9\% &  9\% \\
\hline
   \quad $\Omega$  & -- & 0.066 & 0.021 & 0.012 \\
\hline
\end{tabular}
\end{center}
\caption{Relevant masses and mass differences (in GeV),
   the higgsino fraction of the LSP, the
   most important contributions and the
   resulting $\Omega$ for $m_0=2$ and 3.8~TeV, $m_{1/2}=144$~GeV,
   and $\tan\b=10$ ($A_0=0$, $\mu>0$, $m_t=175$\,GeV).}
\label{tab:largem0}
\end{table}

Another comment is in order. 
Within any of the spectrum codes a change in $m_t$ of the
order of what will be measured at LHC ($\Delta m_t\sim 1$~GeV) induces
large changes in the value of $\mu$ and hence in the LSP mass, 
its higgsino fraction, and the relic density.
The latter can vary by over an order of magnitude within a given code.
This is due to the extreme sensitivity of the running of $m_{H_2}^2$  
to the top Yukawa coupling as explained in Section~2, c.f.\ eq.~(5).
Small changes in the input value of $m_t$ can therefore 
bring approximate agreement between the different codes. 
We emphazise however that this 
only reflects the large theoretical uncertainty in this regime.

\subsection{\mbf Large $m_0$, large $\tan\beta$}

As we increase $\tan\beta$ it becomes increasingly easier to reach
the focus point region. There is also a strong dependence on the value
of the top-quark mass, and typically \isajet\ can find a focus
point behaviour with significantly heavier $\mt$ than the other codes
\cite{Allanach:2004jh}. We consider in more details the case
$\tan\beta=50$ and $m_t=175$~GeV. A value for the relic density in
agreement with WMAP requires $M_1<\mu<M_2$ so that the LSP is a
mixed bino-higgsino state. As one moves very close to the
electroweak symmetry breaking border and $\mu$ drops even below
$M_1$, the higgsino fraction increases rapidly; the relic density
drops below the WMAP range. In what follows we concentrate again
on collider-friendly scenarios with not so heavy neutralinos and charginos.

At large $\tan\b$ and large $m_0$, the main neutralino annihilation channels
are into fermion pairs or into pairs of gauge or Higgs bosons.
Fermion pair production proceeds through  s-channel exchange of Higgs or Z
(the Goldstone component) and is proportionnal to the fermion mass. 
Annihilation into $tt$ is therefore favoured as soon as it becomes 
kinematically accessible.
If not, W-pair production is the dominant channel, proceeding via
t-channel exchange of charginos. Neutralino/chargino co-annihilation channels
can be important as well, but  typically  they are so efficient that
they lead to $\Omega$ below the WMAP range.
The $\mu$ parameter, which determines the neutralino and chargino masses
as well as the $\nt_1\nt_1Z/A$ coupling, also has a significant influence
on the relic density.
In \cite{Allanach:2004xn} it was shown that $\Delta\mu \approx 1$--2\%
could induce shifts of $10\%$ in $\Omega$ for $\tan\beta=50$.
The dependence on $m_{\nt_1}$ or on the pseudoscalar mass is expected
to be weaker; corrections of $50\%$ or larger are necessary
to induce a $10\%$ shift in $\Omega$ \cite{Allanach:2004xn}.

%
\begin{table}[t]\begin{center}
\begin{tabular}{l||cccc}
                & \softsusy\ & \spheno\ & \suspect\ \\
\hline
\hline
   \qquad $\nt_1$    & 135.0 & 148.9 & 146.5 \\
   \qquad $\ch_1$    & 184.0 & 287.0 & 256.0 \\
   \qquad $\nt_2$    & 195.9 & 286.9 & 257.4 \\
   \qquad $\nt_3$    & 212.9 & 502.7 & 324.5 \\
   \qquad $h^0$      & 121.6 & 122.2 & 121.6 \\
   \qquad $A^0$      & 1200  & 1425  & 957 \\
\hline
   $f_H(\nt_1)$         & 30\% & 1.1\% & 4.3\% \\
\hline
   $\nt_1\nt_1\to b\bar b$     &  5\% & 27\% & 44\% \\
   $\nt_1\nt_1\to \tau\tau$    &  --  &  4\% &  6\% \\
   $\nt_1\nt_1\to ZZ$          & 18\% &  7\% &  6\% \\
   $\nt_1\nt_1\to WW$          & 61\% & 29\% & 21\% \\
   $\nt_1\nt_1\to Zh$          &  8\% & 15\% & 10\% \\
   $\nt_1\nt_1\to hh$          &  5\% & 15\% & 10\% \\
\hline
   \qquad $\Omega$ & 0.125 & 18.6 & 2.15 \\
\hline
\end{tabular}
\end{center}
\caption{Relevant masses (in GeV), the higgsino fraction of the LSP,
the most important contributions and the resulting $\Omega$
for $m_0=3450$~GeV, $m_{1/2}=350$~GeV, $\tan\beta=50$, $A_0=0$, $\mu>0$.}
\label{tab:largem0-tb50}
\end{table}

We consider more closely the point $\m0=3450$~GeV, $\mhf=350$~GeV
and $\tan\beta=50$. Table \ref{tab:largem0-tb50} displays the results
for the spectrum and the most important contributions. For this
scenario \softsusy\ gives a result within the WMAP range.
\isajet, however, does not find a solution to the RGEs; the 
relic density of \spheno\ and \suspect\ is orders of magnitude 
above the WMAP bound. 
The reason for the latter is that the $\mu$ parameter in \suspect\ 
and even more in \spheno\ is much larger and one is in a regime of 
a mostly bino LSP --hence no efficient channel for
annihilation is available (c.f.\ the values for $m_{\nt_3}$ 
and $f_H(\nt_1)$ in Table \ref{tab:largem0-tb50}). 
Moreover, \softsusy\ predicts much lighter charginos, which makes 
annihilation into W pairs through chargino exchange more efficient. 
Owing to the huge $\nt_1$--$A^0$ mass difference, 
the influence of the pseudoscalar mass on the relic density is small, 
although there is a spread in the prediction of $\ma$ of several
hundred GeV. This discrepancy in $\ma$ becomes, however, very relevant
in the Higgs funnel region, that is for larger values of $m_{1/2}$.

As was the case in the previous section, 
within any of the spectrum codes a small change in $m_t$ induces
large changes in the value of $\mu$ and hence the relic density, 
which can vary as before by over an order of magnitude within a given code.
Using a different input value for $m_t$ can therefore 
compensate the large discrepancies observed between different
codes.   For example,  a decrease of about 0.5 GeV in $m_t$
brings the results of \suspect\ for both the spectrum and the
relic density, in good agreement with those of \softsusy. 
\softsusy's results of Table \ref{tab:largem0-tb50} can also be 
approximately reproduced with \isajet\ using $m_t=176.36$~GeV. 
Note however that this amounts to extreme fine-tuning.

\subsection{\mbf Varying $A_0$}

Non-zero values of $A_0$ can significantly influence the scalar masses 
as well as the $\mu$ parameter. Roughly speaking, for $A_0<0$
\footnote{Using SLHA conventions, the off-diagonal element of the
   $\{\rm up,\,down\}$-type sfermion mass matrix is 
   $m_{LR}^2=(A_f-\mu\{\cot\b,\tan\b\})m_f$.}
the $\st_1,\sb_1,\stau_1$ masses decrease while $\ma$ and $\mu$ increase.  
For $A_0>0$, the shifts go in the opposite directions.
The pseudoscalar mass is relevant for annihilation processes at 
large $\tan\beta$ where $\nt_1\nt_1\to A^0\to f\bar f$. 
The $\mu$ parameter determines the higgsino fraction of the LSP. 
It also directly influences the mixing in the stau sector and therefore 
the contribution of the $\stau$ coannihilation processes.  
With our convention, $A_0=0$ leads to $A_t<0$ at the weak scale; 
$A_0<0$ increases $|A_t|$ (at the weak scale) thus lowering the $\st_1$ 
mass through i) RG running and ii) a larger $\st_L$--$\st_R$ mixing.
Analogous arguments hold for sbottoms and staus, though here the 
L--R mixing is dominated by $\mu\tan\b$. Also the running of $A_\tau$ is less 
strong, so that $A_\tau$ usually does not change sign with respect to $A_0$. 
The masses and trilinear couplings of the third generation enter in turn 
the running of the Higgs mass parameters, the radiative corrections to 
the Higgs pole masses, and the computation of $\mu$. 

The uncertainties in the masses, estimated as the differences between the codes,
tend to be larger for $A_0\not=0$ as compared to $A_0=0$. Nevertheless the general
picture outlined in the previous sections holds, as the same mechanisms as for 
$A_0=0$ are at work for neutralino (co-)annihilation over most of the parameter 
space. Only when $|A_0|$ becomes large enough to make $\st_1$ very light, 
in fact the NLSP or NNLSP, 
new co-annihilation channels appear associated with a new region of parameter 
space where the relic density is consistent with WMAP.

Let us discuss the cases of moderate and large $\tan\b$ in more detail. 
For $\tan\b=10$, a non-zero value of $A_0$ shifts the contours of
constant light Higgs masses (towards lower values of $\mhf$ for $A_0<0$)  
and moves the position of the stau co-annihilation strips as well as
of the excluded regions (towards higher values of $m_0$ for $A_0<0$)  
as compared to \fig{tb10}. It does however not change the picture 
qualitatively; the WMAP allowed regions are a small bulk region with 
$\nt_1\nt_1$ annihilation and a narrow strip of co-annihilation with staus. 
There is an increase in the differences in $m_{\stau_1}-\mnt{1}$ and hence 
in $\Omega$ between the codes, but the effect is in general not very large.  
The only new feature appears for values of $|A_0|$ large enough to make $\st_1$ 
the (N)NLSP. This case will be illustrated later in this section. 

\begin{figure}[p]
\centerline{\epsfig{file=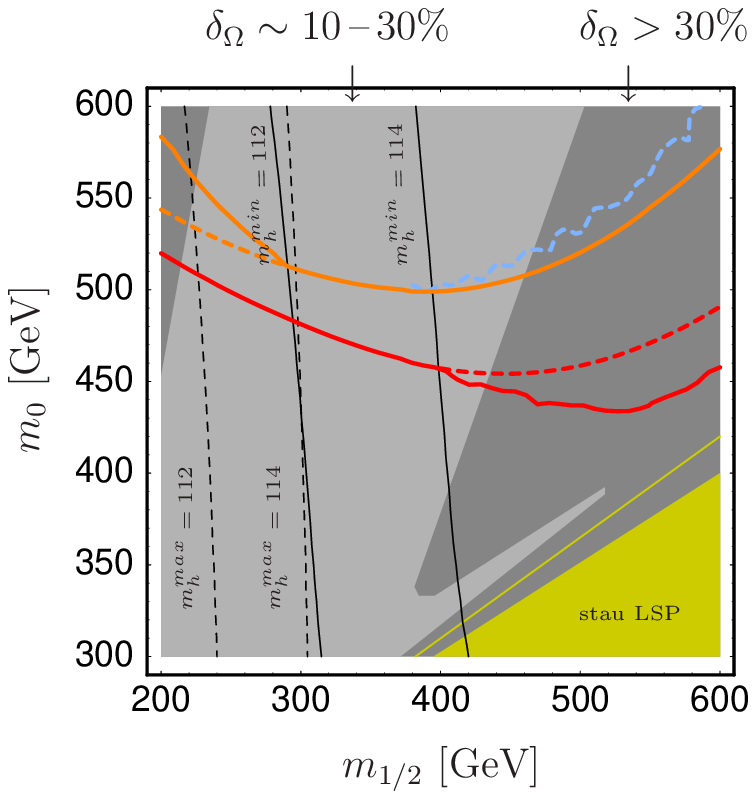, height=8cm}}
\caption{Comparison of results analogous to \fig{tb4050}b ($\tan\beta=50$)
         but for $A_0 = m_{1/2}$.
         The red and orange lines show the variation of the bound
         $\Omega<0.1287$ due to differences in the spectra. 
         The dashed red and orange lines show the situation when 
         only comparing \softsusy\ and \spheno.  
         The gap between the dashed and the full red lines is due to a 
         lighter $\stau_1$ and hence more $\stau$ co-annihilation in \isajet;  
         the gap between the dashed and the full orange lines is due to 
         smaller $\nt_1$ and $A^0$ masses in \isajet. 
         The dashed blue line shows again how the maximal $\Omega$  
         moves when including \suspect.}
\label{fig:tb50_A0mhf}
\end{figure}

For $\tan\beta=50$, we observe larger discrepancies between the codes even for 
moderate values of $A_0$. This is not surprising as the mixing in the $\stau$ 
sector depends on $\mu\tan\beta$ and relatively small shifts in $\mu$ can have 
important effects on the $\stauo$ mass. Moreover,  the pseudoscalar mass
$\ma$ is quite sensitive to $A_0$. While for $A_0=0$ (and small to medium $m_0$)
\isajet, \softsusy, \spheno\ and \suspect\ typically agree on $\ma$ to 1--2\%, 
for $A_0\not=0$ differences of a few per-cent can show up.  
\Fig{tb50_A0mhf} compares the regions of the $m_0$--$\mhf$ plane compatible 
with the upper limit of WMAP analogous to  \fig{tb4050}b but for $A_0=\mhf$.  
As can be seen, the WMAP bound is shifted towards higher values of $m_0$. 
This is because, as mentionned above, the pseudoscalar mass decreases with 
increasing $A_0$, so annihilation channels through Higgs exchange are favoured. 
The Higgs exchange dominates over most of the region of the plot,
however with rather large differences in $\Omega$.
The largest differences are found for $\mhf\gsim 400$~GeV, 
as was the case for $A_0=0$. 
Differences in the pseudoscalar masses increase with increasing $A_0$ 
and $\mhf$, with \softsusy\ and \suspect\ predicting smaller
$\ma$ than \spheno\ and \isajet\ for $A_0=\mhf\gsim 300$~GeV (for smaller values, 
it is \isajet, which predicts the lightest $\ma$). \isajet\ also predicts a 
lighter LSP and a lighter $\stauo$ in the co-annihilation range and hence a 
much lower $\Omega$ as one moves closer to the $\stauo$-LSP boundary.
\suspect\ on the other hand predicts larger $\stauo$ but smaller pseudoscalar  
masses than the other programs. This leads to a larger value for $\Omega$ 
from the \suspect\ spectrum for $A_0=\mhf\gsim 400$~GeV. 
In \fig{tb50_A0mhf}, the dashed blue line shows how the exclusion curve 
correponding to the maximal $\Omega$ is shifted when including \suspect. 
One can conclude that for $\tan\beta=50$ differences between the codes are 
large everywhere, with $\delta\Omega$ exceeding $30\%$ in a large portion 
of parameter space.

%


\begin{table}[p]\begin{center}
\begin{tabular}{l||cccc}
      & \isajet\ & \softsusy\ & \spheno\ & \suspect\ \\
\hline
\hline
   \qquad $\nt_1$    & 161.4 & 164.9 & 164.4 & 164.9 \\
   \qquad $\stau_1$  & 165.4 & 181.5 & 177.4 & 187.5 \\
   \qquad $\ti e_R$  & 406.7 & 406.0 & 406.6 & 408.3 \\
   \qquad $h^0$      & 118.9 & 115.9 & 116.4 & 116.1 \\
   \qquad $A^0$      & 427.3 & 422.0 & 427.6 & 418.0 \\
\hline
   $m_{\stau_1}-\mnt{1}$ & 4.0   & 16.6  &  13.0 & 22.6  \\
   $\ma-2\mnt{1}$        & 104.5 & 92.2  &  98.8 & 88.2\\
\hline
   $\nt_1\nt_1\to b\bar b$     &  3\% & 45\% & 30\% & 66\% \\
   $\nt_1\nt_1\to \tau\tau$    &  --  &  9\% &  7\% & 12\% \\
   $\nt_1\stau_1\to h\tau$     & 21\% & 17\% & 24\% &  7\% \\
   $\nt_1\stau_1\to \g/Z\tau$  & 11\% & 10\% & 14\% &  4\% \\
   $\stau_1\stau_1\to b\bar b$ &  8\% &  6\% &  7\% &  3\% \\
   $\stau_1\stau_1\to hh$      & 28\% &  3\% &  7\% &  --  \\
   $\stau_1\stau_1\to WW,ZZ,\gamma\gamma$   & 15\% &  1\% &  -- &  --  \\
\hline
   \qquad $\Omega$ & 0.017 & 0.107 & 0.081 & 0.136 \\
\hline
\end{tabular}
\end{center}
\caption{Relevant masses and mass differences (in GeV),
the most important contributions and the resulting $\Omega$
for $m_0=376$~GeV, $m_{1/2}=400$~GeV, $A_0=-400$~GeV,
$\tan\beta=50$ and $\mu>0$. 
The higgsino fraction $f_H(\nt_1)$ is 0.8\%.}
\label{tab:A0tanbe50}
\end{table}

%
\begin{figure}[p]
\centerline{\epsfig{file=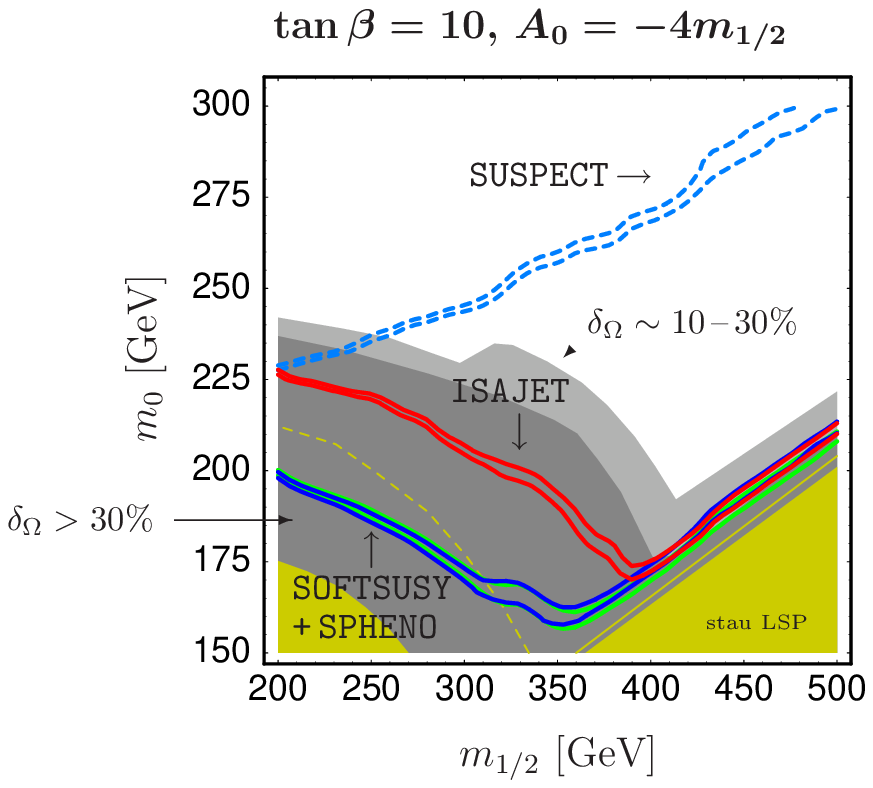, height=8cm}}
\caption{WMAP strips from the four public codes 
         for $A_0=-4m_{1/2}$, $\tan\b=10$,
         $\mu>0$ and $m_t=175$\,GeV. 
         The yellow region in the bottom right corner is excluded 
         due to a $\stau_1$ LSP. 
         In the yellow bottom left region, 
         \softsusy\ and \spheno\ have a $\st_1$ LSP; the yellow dashed 
         line shows the bound of $\st_1$ LSP in \isajet.}
\label{fig:stopcoann}
\end{figure}


\begin{table}[p]\begin{center}
\begin{tabular}{l||cccc}
      & \isajet\ & \softsusy\ & \spheno\ & \suspect\ \\
\hline
\hline
   \qquad $\nt_1$    & 140.8 & 143.2 & 142.5 & 143.0 \\
   \qquad $\stau_1$  & 156.1 & 157.8 & 158.9 & 160.7 \\
   \qquad $\st_1$    & 153.7 & 173.3 & 172.7 & 109.7 \\
   \qquad $h^0$      & 108.8 & 114.1 & 115.6 & 108.3 \\
\hline
   $m_{\stau_1}-\mnt{1}$ &  15.3  & 14.6  & 16.4  &  17.7 \\
   $m_{\st_1}-\mnt{1}$   &  12.9  & 30.1  & 30.2  &  $-33.3$ \\
\hline
   $\nt_1\nt_1\to ee,\mu\mu$   &  --  & 18\% & 16\% &  -- \\
   $\nt_1\nt_1\to \tau\tau$    &  --  & 22\% & 19\% &  -- \\
   $\nt_1\nt_1\to \g/Z\tau$    &  --  & 14\% & 10\% &  -- \\
   $\nt_1\st_1\to th$          & 28\% & 30\% & 36\% &  -- \\
   $\nt_1\st_1\to tg$          &  4\% &  6\% &  7\% &  -- \\
   $\nt_1\st_1\to Zt/Wb$       &  2\% &  4\% &  4\% &  -- \\
   $\st_1\st_1\to gg$          &  4\% &  --  &  --  &  -- \\
   $\st_1\st_1\to gh$          &  2\% &  --  &  --  &  -- \\
   $\st_1\st_1\to hh$          & 57\% &  2\% &  3\% &  -- \\
\hline
   \qquad $\Omega$ & 0.004 & 0.116 & 0.120 & -- \\
\hline
\end{tabular}
\end{center}
\caption{Relevant masses and mass differences (in GeV),
the most important contributions and the resulting $\Omega$
for $m_0=161$~GeV, $m_{1/2}=350$~GeV, $A_0=-1400$~GeV,
$\tan\beta=10$ and $\mu>0$. $f_H(\nt_1)\simeq0.4\%$.}
\label{tab:A0stop}
\end{table}

We have also studied the case $A_0=-\mhf$ at $\tan\b=50$. 
Here the pseudoscalar mass is larger than in the $A_0=0$ case,  
so a relic density in agreement with WMAP requires, especially at large $\mhf$, 
some contribution from co-annihilation processes, in particular with $\stauo$. 
Therefore the value of the relic density is once again very sensitive to the 
$\nt_1$--$\stauo$ mass difference, and discrepancies in $\Omega$ are larger than 
for $A_0=0$. Since for $A_0<0$ we encounter instabilities 
in the scan with \spheno, 
we do not show a plot but examplify this case 
in Table \ref{tab:A0tanbe50} for $m_0=376$~GeV, $\mhf=-A_0=400$~GeV, $\tan\beta=50$ 
and $\mu>0$. Here the discrepancy in the $\stauo$ mass reaches about 10\%,  
meaning that the $\nt_1$--$\stauo$ mass difference varies by more than 100\%, 
thus inducing huge differences in the relic density. The lightest 
$\stauo$ is again obtained with \isajet. We do however find rather good agreement 
between the codes as concerns the boundary of the WMAP region.

A special case is a very large negative $A_0$, such that $\st_1$ becomes
light enough to contribute to co-annihilations. This is the case when the 
$\st_1$ is the NNLSP or even the NLSP. The relic density is then 
very sensitive to the mass difference between $\nt_1$ and $\st_1$.
Since the largest discrepancies between spectrum calculators are usually 
found for the masses of coloured sparticles \cite{Allanach:2003jw},
the predictions for the relic density and for the region compatible with  
WMAP can differ significantly in this case. 
\Fig{stopcoann} shows the WMAP-allowed strips in the $m_0$--$\mhf$ plane 
for $A_0=-4\mhf$, $\tan\b=10$ and $\mu>0$. 
For \isajet, \softsusy\ and \spheno, co-annihilation with stops 
dominates when $m_{1/2}\lsim 350$--400~GeV, while for larger $m_{1/2}$ 
one has mostly stau co-annihilation. For \suspect, $\st_1$ 
co-annihilation dominates over the whole allowed region. 
As expected, the allowed bands are very narrow. They correspond to 
$m_{\st_1}-m_{\nt_1} \approx 20$--30~GeV, typically a much larger mass 
difference than for the case of $\stau$ co-annihilation. 
This is due to the large cross section of $\nt_i\st_1^{}\to th,tg$.
Table~\ref{tab:A0stop} shows the spectrum as well as the most important 
contributions to $\Omega$ for one point, $m_0=161$~GeV, $\mhf=350$~GeV, 
$A_0=-1400$~GeV,  $\tan\beta=10$ and $\mu>0$. 
As one can see, \softsusy\ and \spheno\ agree quite well on the 
$\st_1$ mass 
and hence the relic density, with only few per-cent difference between the 
two programs. In comparison, \isajet\ predicts a lighter $\st_1$ and thus 
a much smaller relic density at a given parameter point. The difference is 
at the level of 10\% for $\mst{1}$ and of ${\cal O}(100\%)$ for $\Omega$. 
Also the boundaries where $\st_1$ becomes the LSP are quite different 
between \softsusy/\spheno\ on the one side and \isajet\ on the other side. 
Part of the discrepancies may come from large logs in the RGEs in \isajet\ 
due to the very large mass splitting of the stops.  
Much larger discrepancies are however found when comparing with \suspect. 
Since \suspect\ does not have the 2-loop RGEs for the squark parameters, 
including $A_t$, it predicts a much lighter $\st_1$ than the other 
three programs. 
For the point of Table~\ref{tab:A0stop}, the difference in $\mst{1}$ 
is about 60~GeV, or 35\%, making $\st_1$ the LSP in the \suspect\ spectrum. 
The large discrepancy between \suspect\ and the other programs can 
be seen clearly in \fig{stopcoann}. Again, the results of \suspect\ are 
reproduced by using only 1-loop RGEs for squark and slepton parameters 
in \softsusy. 
Last but not least notice also that for the parameters of \tab{A0stop}, 
in \softsusy\ and \spheno\ even though $\stau_1$ is the NLSP, 
coannihilation channels with $\st_1$ dominate.

In summary, at very large $A_0<0$ one can get phenomenologically very 
different scenarii for the same mSUGRA point; it is clear that including 
the full two-loop RG running plus a careful treatement of threshold 
corrections is important for a reliable prediction of the relic density.

In this context it is also interesting to compare with the results of 
\cite{Ellis:2004tc}. The `best fit' points in their Fig.~15 are 
\footnote{Note that Ref.~\cite{Ellis:2004tc} uses the opposite sign 
     convention for the trilinear $A$ couplings!}
$m_0=60$~GeV, $m_{1/2}=300$~GeV, $A_0=300$~GeV for $\tan\b=10$ and
$m_0=550$~GeV, $m_{1/2}=500$~GeV, $A_0=1280$~GeV for $\tan\b=50$,
both obtained with $m_t=178$~GeV and $m_b(m_b)=4.25$~GeV \cite{sven}. 
Both are scenarii of stau co-annihilation. 
The relevant masses, mass differences and the resulting values for $\Omega$ 
of Ref.~\cite{Ellis:2004tc} ({\tt SSARD}) are given in Tables~\ref{tab:olive1} 
and \ref{tab:olive2} together with the predictions from \softsusy, \spheno\ 
and \suspect; we leave out \isajet\ where one cannot adjust $m_b(m_b)$.
For the point with $\tan\b=10$, the $\neuto$ and $\stauo$ masses of {\tt SSARD}, 
which has full 2-loop RGEs, are roughly 2\% higher than those of \softsusy\ and 
\spheno. The $\neuto$--$\stauo$mass difference and consequently 
also $\Omega$ lie within the values of \softsusy\ and \spheno.
For the point with $\tan\beta=50$, however, only {\tt SSARD} has a viable 
spectrum with a neutralino LSP,
while the three public codes get a ${\stau_1}$ 
LSP, about 40--60 GeV lighter than the $\stauo$ in {\tt SSARD}. 
Note also the $\sim 10\%$ heavier $\ma$ from {\tt SSARD} as compared 
to the public codes. 
We can recover a similar $\Delta M(\neuto\stauo)$ as \cite{Ellis:2004tc} 
with \softsusy\ and \spheno\ for $A_0=1170$~GeV. 
In this case we get $\mnt{1}=205$--206~GeV, $m_{\stauo}=218$--222~GeV,  
$\ma=500$--510~GeV and $\Omega\simeq 0.098$. However, the fact remains  
that at large $\tan\beta$ (and large $A_0$) there are sizeable differences  
between {\tt SSARD} and the public codes.

\section{Online comparison}

For an easy and user-friendly comparison of SUSY spectrum codes, we have
set up a web application at
\begin{center}
  {\tt http://cern.ch/kraml/comparison/ }
\end{center}
Here the user can input mSUGRA parameter points in a web form.
The value of the top-quark mass is also taken as an input while
$m_b(m_b)$ and $\alpha_s$ are fixed to the values hard-coded in \isajetnn.
The mass spectra are then calculated by the latest versions of 
\isajetnn, \softsusynn, \sphenonn\ and
\suspectnn\ and compared in an output table. 
The corresponding values for $\Omega$, $\d\rho$, $\delta a_\mu$,
$B(b\to s\gamma)$ and $B(b\to s\mu^+\mu^-)$\
are calculated with {\tt micrOMEGAs} and also given in the table.
\softsusynn\ is used with the option of full 2-loop running, 
as in this paper.  
For technical reasons, for the computation of $\Omega$ a `static' 
version of {\tt micrOMEGAs} is used which is limited to (co)annihilation 
channels initiated by $\nt_{1,2,3}$, $\ch_1$, $\ti e_R$, $\ti\mu_R$, 
$\stauo$, and $\st_1$. We have checked that this is largely sufficient 
within mSUGRA.
The webpage is also useful for comparisons with other spectrum
codes and/or programs computing the neutralino relic density.


\begin{table}[t]\begin{center}
\begin{tabular}{l||cccc}
      & {\tt\hphantom{xx}SSARD\hphantom{xx}} & \softsusy\ & \spheno\ & \suspect\ \\
\hline
\hline
   \qquad $\nt_1$    & 119.4 & 117.7 & 117.4 & 117.8 \\
   \qquad $\stau_1$  & 129.1 & 126.1 & 127.2 & 129.5 \\
   \qquad $h^0$      & 113.9 & 111.7 & 112.0 & 111.8 \\
   \qquad $\ma$      & 419.4 & 428.5 & 431.2 & 431.5 \\
\hline
   $m_{\stau_1}-\mnt{1}$ &  9.7  & 8.4 &  9.8  &  11.7 \\
   $\ma-2\mnt{1}$        &  181 & 193 &  196  &  196 \\
\hline
   \qquad $\Omega$ & 0.103 & 0.092 & 0.109 & 0.129 \\
\hline
\end{tabular}
\end{center}
\caption{Relevant masses and mass differences (in GeV) 
   and the resulting $\Omega$ for $m_0=60$~GeV, $m_{1/2}=300$~GeV, 
   $A_0=300$~GeV, $\tan\beta=10$, $\mu>0$, $m_t=178$~GeV and 
   $m_b(m_b)=4.25$~GeV.}
\label{tab:olive1}
\end{table}


\begin{table}[t]\begin{center}
\begin{tabular}{l||cccc}
      & {\tt\hphantom{xx}SSARD\hphantom{xx}} & \softsusy\ & \spheno\ & \suspect\ \\
\hline
\hline
   \qquad $\nt_1$    & 211 & 206 & 205 & 206 \\
   \qquad $\stau_1$  & 226 & 167 & 163 & 185 \\
   \qquad $h^0$      & 117 & 116 & 116 & 116 \\
   \qquad $\ma$      & 553 & 495 & 504 & 490 \\
\hline
   $m_{\stau_1}-\mnt{1}$ &  15  & $-39$ &  $-42$  &  $-21$ \\
\hline
   \qquad $\Omega$ & 0.119 & -- & -- & -- \\
\hline
\end{tabular}
\end{center}
\caption{Relevant masses and mass differences (in GeV) for $m_0=550$~GeV, 
   $m_{1/2}=500$~GeV, $A_0=1280$~GeV, $\tan\beta=50$, $\mu>0$, 
   $m_t=178$~GeV and $m_b(m_b)=4.25$~GeV.} 
\label{tab:olive2}
\end{table}

\section{Conclusions}

We have investigated the impact of uncertainties in SUSY spectrum
computations on the prediction of the neutralino relic density.
To this aim we have compared the results of four public spectrum codes, 
\isajet, \softsusy, \spheno\ and \suspect, in the context of mSUGRA. 
For `moderate', i.e.\ not extreme, values of the model parameters,
we found that the codes in general agree quite well, at the level 
of few percent, for the prediction of the SUSY spectrum. This is 
also true at large $\tan\beta$.

Nevertheless these small discrepancies can have a large impact on the
prediction of the relic density of dark matter. We have studied in
detail the most important scenarios for neutralino (co)annihilation. 
In the bulk region (although largely excluded by the LEP bound on $m_h$), 
predictions are under control, that is uncertainties 
are below the experimental uncertainties of WMAP. 
In the co-annihilation region, however, the uncertainties can easily 
exceed 30\%. Most of this is related to the mass
difference between $\stauo$ and the LSP. 
For this estimate of uncertainties we have used the predictions from 
\isajet, \softsusy\ and \spheno. \suspect, which only has 1-loop RG running 
for the sfermion mass parameters as opposed to full 2-loop RG running in 
the other codes, typically finds a higher $\stauo$ mass. 
To reduce the uncertainty originating from the spectrum
calculation to a level below the experimental uncertainty of
WMAP, one needs a precision in the $\neuto$--$\stauo$ mass difference of 
the level of 1 GeV . 
This corresponds to computing the LSP and NLSP masses 
to per-mille accuracy.
Already it has been shown that going from 2-loop to 3-loop RGE running
\cite{Jack:2003sx} induces corrections of about that level.

Similar arguments hold for scenarios where the neutralinos annihilate 
through pseudoscalar exchange. Typically this means an enhanced coupling 
to the pseudoscalar, that is large $\tan\beta$.
The critical parameter in this case is the $m_A-2m_{\nt_1}$ mass difference.  
Although the codes we compared agree at the level of few percent on the 
pseudoscalar mass, this difference together with the difference in the LSP mass 
can again add up to 30\% or more uncertainty in the relic density. 
To improve the precision of $m_A-2m_{\nt_1}$, one not only needs to go 
to higher orders in the RG running but also  a more precise treatment of 
the Yukawa couplings, especially of $h_b$, is needed eventually including 
the full 2-loop corrections. At this level also a precise treatement 
of $h_\tau$ becomes important.
Notice, however, that the dominant source of uncertainty 
in $m_A$ is still the present error in the extraction of $\mbmb$. 

Models with non-zero $A_0$ have usually similar features w.r.t.\  
the relic density as models with $A_0=0$. The exception is the case of a 
very large $A_0$ where the $\st_1$ becomes light enough to contribute to 
co-annihilations. The existing $\sim 10\%$ uncertainty in the prediction 
of the $\st_1$ mass can then lead to order-of-magnitude discrepancies in the 
prediction of the relic density. In particular, \isajet\ predicts a lighter 
$\st_1$ than \softsusy\ and \spheno, and thus a much lower value for $\Omega$.
The prediciton of \suspect\ for $\mst{1}$ is much below that of the 
other programs. In fact, in the $\st_1$ co-annihilation region of  
\softsusy\ and \spheno\ (but also in the one of \isajet), 
\suspect\ does not provide a viable spectrum due to a $\st_1$ LSP. 
This underlines the importance of including the full
2-loop RG running in the sfermion masses.

The picture is however different for extreme scenarios with very
large $\m0$. These are the most difficult models to handle, and large 
discrepancies in the prediction of the spectrum calculators are found. 
This is especially the case for one of the most important
parameters for the calculation of the relic density, $\mu$, which
determines the masses and higgsino fractions of the neutralinos.
Predictions for $\mu$ can vary by a factor of 2 or more, inducing 
huge order-of-magnitude differences in the relic density.
An improvement of the situation requires in particular a much more 
precise computation of the top Yukawa coupling. A precise measurement 
of the top-quark mass, as addressed in \cite{Heinemeyer:2003ud}, 
would also reduce the uncertainty. 
Owing to the extreme sensitivity of $\mu$ to the exact value of $h_t$ 
near the border of REWSB, we consider this region as very unstable.

We conclude that when using the WMAP bound for constraining mSUGRA models, 
uncertainties from the spectrum computation should be taken into account 
in addition to the experimental uncertainty of $\Omega$. For an estimate 
of the theoretical uncertainties one may use the maximal and minimal 
exclusion curves of different state-of-the-art codes, as we have done 
in this paper. The $\delta\Omega$ obtained this way is comparable
to the one obtained in \cite{Allanach:2004xn} by varying the renormalization 
scale within a given spectrum code. 
Finally, this theoretical uncertainty  should also be combined with
the uncertainty arising from the SM input parameters. 

In parameter regions where $\delta\Omega$ originating from spectrum 
uncertainties is at present larger than the experimental 
uncertainty from WMAP, more precise calculations are certainly desirable 
to improve the reliability of relic density predictions within GUT-scale models. 
Such improvements will be even more important in view of the precision 
envisaged by the PLANCK experiment.

\section*{Acknowledgements}

We thank Ben Allanach, Fawzi Boudjema and Werner Porod for useful discussions.
We also thank Sven Heinemeyer and Keith Olive for providing the detailed 
results of SSARD. 
This work was supported in part by GDRI-ACPP of CNRS and by grants
from the Russian Federal Agency for Science, NS-1685.2003.2. 
The work of S.K. is financed by an APART fellowship of the Austrian 
Academy of Sciences.

\section*{Note added}

After this paper has appeared as a preprint on hep-ph, a new version of 
\suspectnn, {\tt v2.3.4}, was published including the 2-loop RGEs for 
squark and slepton parameters. 
Owing to this improvement, the sfermion masses obtained with \newsuspect\ 
agree well with those of \softsusy, the two programs now being on the 
same level in the implementation of radiative corrections. In particular, 
for the $\st$ co-annihilation point of \tab{A0stop}, \newsuspect\ now gives 
a viable spectrum similar to that of \softsusy\ or \spheno, with 
$\mnt{1}=143$~GeV, $\mst{1}=178$~GeV and $\Omega=0.153$.  
This confirms our observation of the importance of these 2-loop terms.
We note, however, that this does not change the $\d\Omega$ shown in the figures, 
since \suspect\ was not taken into account for the estimate of uncertainties.


\end{document}